\documentclass[11pt,a4paper]{article}
\usepackage[T1]{fontenc}
\usepackage{lmodern,amsmath,amssymb,amsthm,mathtools,booktabs,array,microtype}
\usepackage[margin=29mm]{geometry}
\usepackage{xcolor}
\usepackage{tikz}
\usepackage[colorlinks=true,linkcolor=blue!40!black,citecolor=blue!40!black,urlcolor=blue!45!black]{hyperref}
\newcommand{\paperkeywords}{middle convolution, Katz--Long--Moody functor, marked groups, invariant lattices, Hermitian forms, quantum gates}
\hypersetup{pdftitle={Generalized Middle Convolution and the Construction of Quantum Gates},pdfauthor={Haru Negami},pdfkeywords={\paperkeywords}}
\newtheorem{theorem}{Theorem}[section]
\newtheorem{proposition}[theorem]{Proposition}
\newtheorem{corollary}[theorem]{Corollary}

\newtheorem*{galoiscorollary}{Corollary to Example~\ref{ex:nonsimilar}}
\theoremstyle{definition}
\newtheorem{definition}[theorem]{Definition}
\newtheorem*{definition*}{Definition}
\newtheorem{example}[theorem]{Example}
\newcommand{\C}{\mathbb C}\newcommand{\Z}{\mathbb Z}
\newcommand{\R}{\mathbb R}\newcommand{\Q}{\mathbb Q}
\newcommand{\radic}{\operatorname{rad}}
\newcommand{\im}{\operatorname{im}}\newcommand{\diag}{\operatorname{diag}}
\newcommand{\rk}{\operatorname{rank}}\newcommand{\In}{\operatorname{In}}
\newcommand{\KLM}{\mathrm{KLM}}
\newcommand{\doi}[1]{\href{https://doi.org/#1}{doi:\nolinkurl{#1}}}
\newcommand{\arxiv}[1]{\href{https://arxiv.org/abs/#1}{arXiv:#1}}
\numberwithin{equation}{section}
\title{Generalized Middle Convolution\\and the Construction of Quantum Gates}
\author{Haru Negami\\[0.35em]
  {\normalsize Japan Agency for Marine-Earth Science and Technology (JAMSTEC)}\\
  {\small\href{mailto:negamiharu@gmail.com}{\nolinkurl{negamiharu@gmail.com}}}}
\date{December 2026}
\begin{document}
\maketitle
\begin{abstract}
Given a monodromy representation on a punctured sphere, middle convolution
produces a new representation of the same fundamental group. For Fuchsian
systems satisfying suitable hypotheses, this operation is realized by an
integral transformation of solutions.
We give an algebraic formulation of middle convolution for representations
of marked groups, allowing the input and output groups to differ. To
compare multiplicative middle convolution with its additive and
$q$-difference counterparts, we describe the middle space as a quotient
by the radical of a Hermitian form. For self-adjoint coefficients and
real, nonexceptional parameters, a single radical formula recovers the
middle quotients of additive middle convolution and of the
$q$-middle convolution of Sakai and Yamaguchi.
Given a positive definite input representation with a full invariant
lattice, we determine the largest input subrepresentation whose middle
convolution satisfies prescribed output relations. Orthogonal projection
of the full middle lattice gives the invariant lattice on this summand.
For every $k\geq 1$, we construct from a cyclic character a family of
meridians generating a dense subgroup of $U(2^k)$. In a separate
two-qubit example, the relation $T_1^4=I$ selects a positive definite
quotient on which two gates of orders four and eight generate a dense
subgroup of $PU(4)$ after passage to projective classes.
For a rank-two cyclotomic example, we show that the largest expansion
by gate words grows exponentially with their length bound in the
canonical embedding, whose Euclidean norm is used in lattice cryptography.
\end{abstract}
\begin{quotation}
\end{quotation}

\section{Introduction}
In earlier work \cite{NegGates}, we proposed using the Katz--Long--Moody (KLM) functor to construct
quantum gates from braid group representations.
Applying the functor to a character twist of the Burau representation
gives a family of unitary operators in a common orthonormal basis.
We call this approach the \emph{middle convolution gate construction},
or \emph{MC gate construction} for short (Definition~\ref{def:gate}).
Haraoka gives an integral transformation of solutions of KZ equations
and shows, under the spectral hypotheses of his comparison theorem,
that the monodromy of the transformed system is obtained by
multiplicative middle convolution \cite[eq.~(3.2), Def.~5.1 and Thm.~5.5]{HarKZ}.
With compatible conventions for analytic continuation along loops, the KLM
construction of Hiroe and Negami \cite{HN} agrees with this operation on
the pure braid subgroup $P_{n+1}$ \cite[Thm.~6, Cor.~7 and \S4.2]{NegMC}.

Throughout, a group is \emph{marked} if an ordered tuple of its
generators is specified. In the KZ setting these generators are
\emph{meridians} in the convolution variable. The moving point follows
a path from the base point to a small circle around one puncture, goes
once around the circle, and returns along the same path. In our group
presentations the specified generators play the role of these loops.

Two questions motivate this paper.
\begin{enumerate}
\item Can middle convolution be extended to other marked monodromy
representations in a way that preserves unitarity, and when can the
resulting representation be realized by an integral transformation of
solutions of a complex linear differential equation?
\item Can one give a general algebraic construction of quantum gates
that specifies which relations the gates satisfy, which lattice they
preserve, and how they act in one computational basis?
\end{enumerate}

We give an algebraic criterion for the prescribed relations and
identify integral transforms in specific differential systems. A general
criterion for analytic realization remains open.

Beukers and Heckman gave a criterion for an irreducible generalized
hypergeometric equation with rational parameters to have finite monodromy
\cite[Thm.~4.8]{BH}.
Motivated by their proof, we used an invariant lattice together with
the signatures of invariant Hermitian forms at all embeddings of the
coefficient field to determine when middle convolution at a root of unity
preserves finite monodromy \cite[Lem.~2.9 and Thm.~2.10]{NegFinite}. An isometry of a Euclidean
lattice has only finitely many choices for the image of each basis vector.
Over a number field, positive definite invariant forms at all archimedean
places bound the orbits of the lattice generators in the diagonally
embedded lattice, so the image is finite. Radchenko and Rodriguez
Villegas use this argument for rank-four Goursat systems
\cite[\S\S4.3 and~5]{RRV}. Ghate and Venkataramana give the corresponding
finiteness criterion for irreducible integral unitary groups over CM
fields \cite{GV}.

Theorem~\ref{th:ideal} constructs the output lattice to which this
finiteness argument applies. The lattice also constrains the image
at finite primes. Every group element has integral trace, and the
action on the lattice can be reduced modulo any prime ideal. For example, a trace
obstruction at the prime above $2$ rules out a common integral lattice
for the Hadamard and $\mathsf T$ gates (Proposition~\ref{prop:dyadic}).
In our two-qubit example, the conjugate form of signature $(2,2)$ and
absolute irreducibility imply that the image is infinite. Reduction
modulo $17$ certifies that an adjoint orbit spans $\mathfrak{sl}_4$,
a step in proving projective density (Proposition~\ref{prop:fourclosure}).

For the gate construction, the invariant Hermitian form need only be
positive definite at the chosen complex embedding. If the output space
has dimension $2^k$, choose an orthonormal basis for this form and label
its vectors by $k$-bit strings. In this computational basis each marked
generator acts as a unitary gate that maps the output lattice onto itself.
Since the lattice spans the output space, two gate sequences give the
same unitary operator exactly when they agree on every lattice vector.

For KZ-type equations satisfying the comparison hypotheses above,
a full lattice $A$ preserved by the input monodromy determines an
invariant lattice for the output once the convolution parameter is fixed.
More precisely, suppose $\lambda\ne1$ is a unit in the ring of integers
of the coefficient field. The image of $A^{\oplus n}$ in the middle
quotient, with one copy of $A$ for each peripheral generator, is a full
lattice preserved by the output monodromy (Section~\ref{sec:quot}).
The KLM construction computes this monodromy from the action of the
input group on $A$. When $|\lambda|=1$, fixing an invariant positive
definite Hermitian form on the input and a square root of $\lambda$
also determines the induced Hermitian form on the quotient, and hence
its Hermitian matrix in any chosen basis
(Definition~\ref{def:klmform} and Theorem~\ref{th:ideal}).

We use this passage from one invariant lattice to another to give a
purely algebraic extension of middle convolution for representations of
marked groups with prescribed output relations. For an invariant
Hermitian form $H$ on the ambient convolution space $W$ with lattice
$\Lambda$, the middle space and its lattice
(Definitions~\ref{def:middle} and~\ref{def:middlelattice}) are
\[
       W/\ker H,\qquad \Lambda/(\Lambda\cap\ker H).
\]
The form determines the subspace by which we divide. Every group
acting on $\Lambda$ by isometries of $H$ therefore acts on this quotient
lattice. A coefficient operator $F$ satisfying $HF=F^*H$ also preserves
$\ker H$. For example, in the $q$-difference case
Theorem~\ref{th:q} identifies the subspace $\mathcal K+\mathcal L$ in
the definition of Sakai and Yamaguchi with the radical of the same
Hermitian form used for additive convolution, at $c=1-q^{\lambda_q}$, where $\lambda_q$ is the exponent in the $q$-convolution parameter.
The induced form is nondegenerate and is positive definite precisely
when its negative index is zero. In the multiplicative
construction the input and output groups may be any two quotients of
the free group on the chosen marking. Taking the auxiliary symmetry group
to be trivial gives a representation of the output group itself
(Theorem~\ref{th:transfer}). For the KLM coefficient matrices, the
output is ordinary KLM on a summand selected by the relations.
From the marked presentation we determine both the ideal that selects
this summand and the image of the projected input lattice in the quotient. In an absolutely irreducible
rank-two output, projection and intersection give Hermitian lattices
that are not even similar (Example~\ref{ex:nonsimilar}).

\subsection*{Related work}
\paragraph{A. Middle convolution and integral transformations.}
Katz develops middle convolution for local systems on a punctured
line \cite[Chap.~2]{Katz}. Dettweiler and Reiter define multiplicative
and additive middle convolution on tuples of matrices
\cite[Defs.~2.2, 2.5, A.2 and A.6]{DR00}. In the multiplicative case
they construct an invariant pairing that annihilates the sum of the
local and infinity kernels and give its Hermitian version
\cite[Thm.~5.7 and Cors.~5.8--5.9]{DR00}. Their later paper proves
that multiplicative middle convolution preserves a Hermitian form when
$\lambda$ is a root of unity and gives the additive block construction
\cite[Thm.~2.4(v) and \S5.1]{DRpre}. Under the hypotheses
of their comparison theorem, the monodromy of the transformed Fuchsian
system is the multiplicative middle convolution of the original
monodromy \cite[Def.~3.1 and Thm.~4.7(iii)]{DR}. Haraoka defines middle convolution along a chosen coordinate for
completely integrable logarithmic systems on hyperplane complements
and gives an integral representation of solutions of the transformed
system \cite[Def.~2.2 and Thm.~2.2]{Har}. Oshima studies convolution
in a chosen direction of a KZ-type system \cite[\S5]{Osh}. Sakai--Yamaguchi and Arai--Takemura give
$q$-difference analogues \cite{SY,AT}. For the corresponding $q$-integral transformations in hypergeometric examples see \cite{AT23}. Hiroe--Negami identify KLM
with a homological middle image \cite{HN}. For an input preserving a
nondegenerate Hermitian form and $|\lambda|=1$, $\lambda\ne1$, we proved
in \cite[Thm.~9]{NegMC} that the radical of the ambient KLM Hermitian
form is exactly the sum of the local and infinity kernels. Its signature
is computed from the eigenangles in \cite{Neg}. For parabolic cohomology, Dettweiler and Wewers
compute the signature from local eigenangles and give an explicit
formula for the pairing \cite[Thms.~2.3 and~2.6]{DW}.
Dettweiler and Sabbah determine how Hodge data change under middle
convolution \cite{DS,DSerr}. Arinkin identifies Kummer middle convolution with
a restriction of the Katz--Radon transform \cite[\S6.3]{Ari}.
Yamakawa uses Harnad duality, and Hiroe extends additive convolution
to Lie algebra representations \cite{Yam,Hir}. In our construction the radical determines the quotient, and the input
lattice determines its integral structure. Both are defined even for a
marked continuation space with no specified integral kernel.

\paragraph{B. Representations with different acting groups.}
Souli\'e constructs Long--Moody functors for families of groups
\cite{Sou}, and Crawley-Boevey and Shaw realize multiplicative
middle convolution through reflection functors for multiplicative
preprojective algebras \cite{CBS}. Boalch constructs finite braid orbits in $\mathrm{SL}_2(\C)$ from
triples of reflections generating finite complex reflection groups in
dimension three \cite{Boa}. For the middle convolution approach to
finite braid orbits of reflection tuples, see \cite{DRJ,Vay}.
A classical analytic comparison is provided by the Gauss equations
with dihedral monodromy \cite{Vid}. We fix ordered generators for the input and output groups and ask which
input identities ensure that the prescribed output group acts on the
middle lattice.

\paragraph{C. Algebraic constructions of quantum gates.}
The twisted Burau construction in \cite{NegGates} gives generators
and their simultaneous realization for one family of KLM inputs.
Freedman, Larsen and Wang determine closures of Jones--Wenzl braid
images for specified levels and strand numbers \cite[Thm.~0.1]{FLW}. For analytic context, Leksin realizes Burau and
Gassner representations by Jordan--Pochhammer monodromy \cite{Lek}.
Giles and Selinger characterize exact Clifford+$\mathsf T$ synthesis with
ancillas by a coefficient ring \cite{GS}; Kliuchnikov and
Sch\"onnenbeck use Hermitian lattices in circuit synthesis \cite{KS}.
In our two-qubit example the relation $T_1^4=I$ selects the
positive quotient on which two integral meridians generate a
projectively dense group. A cyclic family gives dense unitary images
in every qubit dimension.
In view of the role of the canonical embedding in ring-LWE \cite{LPR}
and recent quantum algorithms for low-rank cyclotomic module lattices
\cite{AnthLattices}, we also ask how much a gate word can change the
Euclidean length of a perturbation across all embeddings.

\subsection*{Main results}
\paragraph{Theorem A-1 (the same marked group).}
We determine the input summand selected by the relations and compute
the resulting action, invariant lattice, and Hermitian signature
(Theorems~\ref{th:descent} and~\ref{th:ideal}).
For cyclic orders and graph products the criterion is explicit
(Theorems~\ref{th:cyclic} and~\ref{th:graph}). Proposition~\ref{prop:artin}
treats alternating relations on character inputs.

\paragraph{Theorem A-2 (comparison with known middle convolutions).}
For self-adjoint coefficients and real parameters, one Hermitian form
recovers the additive, KZ, and $q$-difference middle quotients and
arises from the quadratic term of the multiplicative KLM form
(Theorems~\ref{th:scalar}, \ref{th:dir}, \ref{th:q}, and~\ref{th:tangent}).

\paragraph{Example A-3 (elliptic character functions).}
On a cyclic cover of an elliptic curve, the inverse-character sum
$f_a(-z)$ realizes the output continuation representation
(Proposition~\ref{th:ellipticchars}).

\paragraph{Theorem B-1 (different marked groups).}
We characterize the largest input summand whose middle representation
satisfies the relations of a prescribed output group
(Theorem~\ref{th:transfer}).

\paragraph{Analytic realization B-2 (cyclic to dihedral).}
An Euler integral realizes the passage from a cyclic character to the
reflection representation of a dihedral group
(Example~\ref{ex:dihedral} and Proposition~\ref{prop:dihedralperiod}).

\paragraph{Theorem C (two finite-order gates and a dense family).}
The relation $T_1^4=I$ selects a positive two-qubit quotient on which two
gates of orders four and eight generate a projectively dense group, and cyclic inputs
give dense subgroups of $U(2^k)$ for every $k\ge1$
(Example~\ref{ex:fourgate}, Proposition~\ref{prop:fourclosure}, and
Theorem~\ref{th:densegates}).
For both lattices in Example~\ref{ex:nonsimilar}, we bound Euclidean expansion
in the canonical embedding by gate-word length (Proposition~\ref{prop:canonical-distortion})
and exhibit exponential growth at an indefinite place.

\paragraph{Organization of the paper.}
Section~\ref{sec:hermitian} defines the Hermitian quotient and computes
its signature. Section~\ref{sec:kz-difference} applies it to KZ and
$q$-difference systems. Example~\ref{ex:local-residue-circuit} gives a
two-qubit circuit that estimates a residue from local monodromy.
Section~\ref{sec:tangent} recovers the
additive form from the quadratic term of the multiplicative KLM form.
Section~\ref{sec:integral} studies quotient lattices and the integral
span of Pochhammer cycles, represented on the punctured plane by the
contours used in Euler transforms. Section~\ref{sec:quot} determines the output
representation and lattice for marked quotients, and Section~\ref{sec:graph}
treats commutation and Artin relations among the marked generators. Section~\ref{sec:transfer}
allows different input and output relations and applies the construction
to quantum gates. Section~\ref{sec:analytic} gives elliptic, dihedral,
and Appell examples. Section~\ref{sec:questions} discusses open questions
on analytic realizations and discrete Painlev\'e equations.

\section{The Hermitian quotient and its coefficients}\label{sec:hermitian}
All vector spaces in Sections~\ref{sec:hermitian}--\ref{sec:tangent} are finite-dimensional complex Hilbert
spaces. All adjoints are taken with respect to the given inner products. A Hermitian operator
$H$ represents the form $(x,y)\mapsto x^*Hy$ in orthonormal coordinates.
Write $\In(H)=(n_+(H),n_-(H),n_0(H))$, counting multiplicities.

\begin{definition}[Hermitian middle space and coefficient operators]\label{def:middle}
Let $U:V\to W$ be linear, let $D=D^*$ on $W$, and let $C=C^*$ be
invertible on $V$. Put
\begin{equation}\label{eq:data}
 S=U^*D,\quad T=SU,\quad L=I-UC^{-1}S,\quad
 H=DL=D-S^*C^{-1}S .
\end{equation}
We write $H_C$ for $H$ when we wish to display the operator parameter $C$.
The middle space of these data is $M=W/\radic H$, equipped with the
induced form. For each endomorphism $\Phi$ of $W$ satisfying $D\Phi=\Phi^*D$,
we take $X_\Phi=\Phi L$ as a coefficient operator.
An isometry from these data to $(V',W',U',C',D')$ is a pair of maps
$g_V:V\to V'$ and $g_W:W\to W'$, unitary with respect to the given
positive definite Hilbert space inner products, such that
\[
 g_VC=C'g_V,\qquad g_WD=D'g_W,\qquad g_WU=U'g_V.
\]
\end{definition}
These identities give $g_W^*H'g_W=H$, so $g_W$ induces an isometry
between the middle spaces for their induced Hermitian forms, which may
be indefinite.

\begin{theorem}\label{th:general}
The radical and inertia are given by
\begin{align}
 \radic H&=\ker D\ \oplus\ U\ker(C-T),\label{eq:rad}\\
 \In(H)+\In(C)&=\In(D)+\In(C-T).\label{eq:inert}
\end{align}
The restriction of $U$ to $\ker(C-T)$ is injective. Each $X_\Phi$
induces an endomorphism of $M$ that is self-adjoint with respect to
the induced nondegenerate form. If $C>0$, the induced form is positive definite if and only if
$D\geq0$ and $C-T\geq0$.
\end{theorem}
\begin{proof}
If $Hw=0$, take $u=C^{-1}Sw$. Then $w-Uu\in\ker D$, hence
$Sw=Tu$ and $(C-T)u=0$. Conversely $H$ kills $\ker D$, and
$HUu=DUC^{-1}(C-T)u$. If $Uu\in\ker D$ and $(C-T)u=0$,
then $Cu=Tu=0$, so $u=0$. This proves \eqref{eq:rad}, including the
injectivity assertion.

For the inertia identity, Haynsworth's formula applies with invertible
pivot $C$ and Schur complement $H$ \cite[Thm.~1, p.~75]{Hay}. The
other congruence uses $S=U^*D$ and holds even when $D$ is singular.
Consider the form on $W\oplus V$ represented by
\[
 \begin{pmatrix}D&S^*\\ S&C\end{pmatrix}.
\]
Its value on $(w,v)$ can be written in either of the two ways
\begin{align*}
 (w+Uv)^*D(w+Uv)+v^*(C-T)v,\qquad
 w^*Hw+(v+C^{-1}Sw)^*C(v+C^{-1}Sw).
\end{align*}
The changes of variables are invertible, giving \eqref{eq:inert}.
For the coefficient operators, $H=L^*D$ and
$HX_\Phi=L^*D\Phi L$ is Hermitian. Thus $X_\Phi$ preserves $\ker H$ and
is self-adjoint on the quotient. When $C>0$, the negative component
of \eqref{eq:inert} is a sum of two nonnegative integers, which gives
the criterion.
\end{proof}

Isometries of the data intertwine $H,L$ with $H',L'$ and send $X_\Phi$ to
$X'_{g_W\Phi g_W^{-1}}$. Hence a symmetry group of the data acts by
isometries on $M$. It may permute the chosen coefficients. A
permutation of labels alone becomes an operator on a fixed fiber when
one supplies a compatible identification with the relabeled data.

\subsection{Dependence of the signature on the scalar parameter}
\begin{definition}[The Hermitian form for a scalar parameter]\label{def:scalar}
Take $W=V^m$, $Uu=(u,\ldots,u)$, $D=\diag(B_1,\ldots,B_m)$
with $B_i=B_i^*$, and $C=cI$ for $c\in\R^\times$. Write $E_i:V\to W$
for inclusion into the $i$th summand. For $w=(w_1,\ldots,w_m)\in W$,
write $S(w)=\sum_i B_iw_i$, so $S=\sum_i B_iE_i^*$ and
$T=SU=\sum_iB_i$. For the scalar parameter $c$ we write
$H_c=H_{cI}=D-c^{-1}S^*S$. Its $(i,j)$ block is
\begin{equation}\label{eq:blocks}
 (H_c)_{ij}=\delta_{ij}B_i-c^{-1}B_iB_j.
\end{equation}
The corresponding quotient is $M_c=W/\ker H_c$.
\end{definition}
The factors $B_i$ and $B_j$ need not commute. Taking the adjoint
reverses their order and gives the $(j,i)$ block.

\begin{theorem}\label{th:scalar}
Put $d=\dim V$ and $m_c=\dim\ker(cI-T)$. Then
\begin{align}
 \ker H_c&=\left(\bigoplus_{i=1}^{m}E_i(\ker B_i)\right)
              \oplus U\bigl(\ker(cI-T)\bigr),
 &\dim M_c&=\sum_i\rk B_i-m_c .\label{eq:middle}\\
 n_\pm(H_c)&=\sum_i n_\pm(B_i)+n_\pm(cI-T)-n_\pm(cI).
 \label{eq:sig}
\end{align}
The signature of the form on $M_c$ is the ordered
pair $(n_+(H_c),n_-(H_c))$. If $c>0$ its negative index is
$\sum_i n_-(B_i)+\#\{\tau\in\operatorname{Spec}(T):\tau>c\}$.
If $c<0$ its negative index is
$\sum_i n_-(B_i)-\#\{\tau\in\operatorname{Spec}(T):\tau\le c\}$.
All spectral counts include multiplicities.
\end{theorem}
\begin{proof}
Set $C=cI$ in \eqref{eq:rad} and \eqref{eq:inert}. The inertia
calculation uses Haynsworth's Schur complement formula
\cite[Thm.~1]{Hay}, as in the multiplicative KLM calculation
\cite[eq.~(7)]{Neg}. For $c<0$, subtracting $n_-(cI)=d$ from
the count of eigenvalues of $T$ above $c$ gives the negative of the
complementary count.
\end{proof}

At a nonzero eigenvalue $c_0$ of $T$, restrict the derivative of $H_c$
to the additional kernel $U\bigl(\ker(c_0I-T)\bigr)$, as in
\cite[Lem.~6.5(ii) and Prop.~6.8]{Neg}. Here the crossing form is
positive, since for $Tu=c_0u$,
\[
 (Uu)^*\left.\frac{dH_c}{dc}\right|_{c_0}Uu
     =c_0^{-2}\|SUu\|^2=\|u\|^2.
\]
Thus, as $c$ increases through $c_0$, $m_{c_0}$ directions pass from
negative to positive, and disappear in the middle quotient at $c=c_0$. The local kernel $\ker D$ is present on both sides. The parameter
zero is excluded throughout.

With $P_i=E_iE_i^*$, the additive convolution row of
\cite[Def.~A.2]{DR00} at parameter $-c$ is $R_i=-cX_{P_i}$. More
explicitly, if $Q_i=S-cE_i^*$, then
\begin{equation}\label{eq:row}
 R_i=E_iQ_i,\qquad H_c R_i=-c^{-1}Q_i^* B_i Q_i .
\end{equation}
Thus $H_c$ determines the middle kernel and simultaneously
symmetrizes all the rows. A linear map $f$ satisfying $fB_i=B'_if$
for every $i$ induces a map between the middle spaces. Indeed,
$f^{\oplus m}$ carries each summand of \eqref{eq:middle} into its
counterpart and intertwines the rows. This construction preserves
compositions and identities. If the induced form on $M_c$ is positive
definite, $M_c$ can serve as the input Hilbert space for a further
convolution, to which \eqref{eq:sig} again applies.

For a nonscalar parameter, the formula $X_{P_i}=E_i(E_i^*-C^{-1}S)$
still holds when $[C,B_i]\ne0$. Thus the construction in
Theorem~\ref{th:general} also specifies which order to use for the
matrix parameter. For example, take $m=2$,
$B_1=\diag(1,0)$, $B_2=\diag(0,1)$ and
$C=\left(\begin{smallmatrix}2&1\\1&2\end{smallmatrix}\right)$.
Then $T=I$, both commutators with $C$ are nonzero, and $C-T$ is
positive semidefinite of rank one. The output is a positive Hermitian
line. Its dimension follows from $\rk D=2$ and $\dim\ker(C-T)=1$.

For $C>0$ in this block model one can identify the whole tuple over
$\C$. Put $P=C^{1/2}$, $A_i=P^{-1}B_iP^{-1}$ and $F=I_m\otimes P$.
Writing the input tuple in parentheses, direct multiplication gives
\[
 H_C(B)=F^*H_1(A)F,\qquad
 F X_{P_i}(B,C) F^{-1}=X_{P_i}(A,I).
\]
Thus, over $\C$, this change of input tuple reduces every positive
definite matrix parameter to the scalar parameter $1$. Whether this identification carries a specified integral
lattice onto another one is an arithmetic question.

\section{Middle convolution of KZ and difference systems}\label{sec:kz-difference}
\begin{definition*}[KZ-type equation]
Let $V$ be a finite-dimensional complex vector space. On the configuration
space of ordered, pairwise distinct points $x_0,\ldots,x_m\in\C$, a
Knizhnik--Zamolodchikov type (KZ-type) equation is the system
$\nabla u=0$ for a flat logarithmic connection of the form
\[
 \nabla=d-\sum_{0\le i<j\le m} A_{ij}\,d\log(x_i-x_j).
\]
Here $A_{ij}\in\operatorname{End}_{\C}(V)$ is the constant residue
matrix along $x_i=x_j$, and we write $A_{ji}=A_{ij}$.
The integrability condition $\nabla^2=0$ is equivalent to the
infinitesimal braid relations \cite[eqs.~(3.1)--(3.3)]{Osh}
\begin{equation}\label{eq:kohno}
 [A_{ij},A_{k\ell}]=0,\qquad [A_{ij},A_{ik}+A_{jk}]=0
\end{equation}
for four, respectively three, distinct indices.
\end{definition*}

\subsection{Middle convolution of a KZ system with respect to one variable}
\begin{definition}[Residues of convolution with respect to $x_0$]\label{def:directional}
For the Hermitian construction, assume that the residues of the KZ-type
equation are self-adjoint with respect to a common positive definite
inner product on $V$, so $A_{ij}^*=A_{ij}$ for all $i<j$.
Choose $x_0$ as the convolution variable, put $B_i=A_{0i}$ for
$1\le i\le m$, and fix a nonzero real additive parameter $\mu=-c$.
In the convention of \cite[\S5]{Osh} the enlarged
residue at $(0,i)$ is $R_i$ of \eqref{eq:row}. The enlarged transverse
residue at $(i,j)$ is
\begin{equation}\label{eq:trans}
 R_{ij}=I_m\otimes A_{ij}
       +(E_i B_j-E_j B_i)(E_i^*-E_j^*).
\end{equation}
\end{definition}

We index only the finite points. For a fixed variable, its residue
at infinity is the negative of the sum of its finite residues.

\begin{theorem}\label{th:dir}
For the data of Definition~\ref{def:directional}, middle convolution with respect to $x_0$ has underlying space $M_{-\mu}$, and every transformed residue is self-adjoint for its
induced form. In particular, if $\mu<0$ the form on the quotient is
positive definite exactly when $A_{0i}\geq0$ for each $i$ and
$\sum_i A_{0i}\le -\mu I$. The same statements hold for any choice of variable after relabeling the indices.
\end{theorem}
\begin{proof}
Put $A_{0\infty}=-\sum_iA_{0i}$. The kernel defining the quotient
consists of the diagonal summand $U\ker(A_{0\infty}-\mu I)$ and the
local coordinate kernels. Haraoka proves that both subspaces are
invariant under the transverse equations \cite[Prop.~2.2]{Har}.
Here their sum is the radical in \eqref{eq:middle}.
Equation~\eqref{eq:row} treats the residues $R_i$.

For $R=R_{ij}$ first compute $DR$. Its diagonal blocks at $i,j$ are
$B_i(A_{ij}+B_j)$ and $B_j(A_{ij}+B_i)$, both Hermitian by
\eqref{eq:kohno}. The other diagonal blocks are $B_k A_{ij}$ with
$[B_k,A_{ij}]=0$. The off-diagonal blocks are $-B_iB_j$ and
$-B_jB_i$, which are adjoints. Thus $DR=R^*D$. Moreover
$RU=UA_{ij}$ by \eqref{eq:trans}. Taking adjoints and using
$S=U^*D$ gives $SR=A_{ij}S$. Consequently
\[
 H_cR=DR-c^{-1}S^* A_{ij}S
\]
is Hermitian. Hence each residue induces a self-adjoint operator
on the quotient, as does every real linear combination of residues. The positivity criterion follows from
Theorem~\ref{th:scalar}.
\end{proof}

The form here makes the transformed residues self-adjoint. Parallel
transport for the logarithmic connection involves the complex
differentials $d\log(x_i-x_j)$ as well. The coefficient form and
a monodromy-invariant form can have different signatures, as in
Section~\ref{sec:appell}.

\subsection{Two normalizations of \texorpdfstring{$q$}{q}-middle convolution}
Use the index set $0,\ldots,N$ and write $\eta=q^{\lambda_q}$ for the chosen
value of the multiplicative parameter. Sakai and Yamaguchi denote its
exponent by $\lambda$ \cite[Def.~1.5, eq.~(20), p.~1981]{SY}. In their notation for the coefficients,
\[
 B(x)=B_\infty+\sum_{i=1}^N\frac{B_i}{1-x/b_i},
 \qquad B_0=I-B_\infty-\sum_{i=1}^N B_i .
\]
Thus $T=I-B_\infty$. Suppose all $B_0,\ldots,B_N$ are self-adjoint
and $\eta\in\R^\times\setminus\{1\}$. Put $c=1-\eta$, take $S,U$ on
$W=V^{N+1}$, and use the corresponding coordinate inclusions $E_i$.

\begin{theorem}\label{th:q}
The Sakai--Yamaguchi quotient and the Arai--Takemura quotient for
these data both have underlying Hermitian space $M_{1-\eta}$. In the
first normalization the induced coefficients are represented by
\[
 F_i=E_i(S-cE_i^*)\quad(1\leq i\leq N),\qquad F_\infty=I-US .
\]
In the second they are represented by
\[
 G_i=\eta^{-1}E_i(S-cE_i^*)\quad(0\leq i\leq N).
\]
All these coefficients are self-adjoint for the induced form. In
particular, for $0<\eta<1$ this form is positive definite precisely when
$B_i\ge0$ for every $0\le i\le N$ and $B_\infty\ge \eta I$.
\end{theorem}
\begin{proof}
We use the coefficient matrices and quotient of Sakai and Yamaguchi
\cite[Defs.~1.5--1.6, p.~1981]{SY} and the reformulation by
Arai and Takemura \cite[Def.~3.1, Prop.~3.10, Def.~3.11 and eq.~(5.3)]{AT}.
The published Definition~1.6 prints $\ker A_i$ in the formula for $\mathcal K$.
We use $\mathcal K=\bigoplus_{i=0}^N(\ker B_i)$ as in Definition~1.3,
eq.~(7) of the arXiv version of \cite{SY}, with $B_i$ as in the published
eq.~(20). The additional kernel is $\ker(US-cI)$:
if $USw=cw$ then $w=Uu$ with $Tu=cu$, and the converse follows from
$SU=T$. This gives the radical \eqref{eq:middle}.
The row expressions are the defining coefficient blocks, with
$1-\eta^{-1}=-\eta^{-1}c$ in the second normalization. Identity~\eqref{eq:row}
covers the finite rows. For the remaining matrix, use
\[
 H_cUS=S^*(I-c^{-1}T)S=(H_cUS)^*.
\]
Finally, $T\le (1-\eta)I$ is equivalent to $B_\infty\ge \eta I$.
\end{proof}

In both normalizations each coefficient $F$ satisfies
$H_cF=F^*H_c$. Thus $F$ sends $\ker H_c=\mathcal K+\mathcal L$ into
itself, and the algebra generated by the coefficients in either
normalization acts on the same middle space. An invariant metric for the $q$-difference system itself
would have to satisfy $B(x)^*h(qx)B(x)=h(x)$ on a specified domain.
This additional condition relates the metrics on the fibers at $x$ and
$qx$.

\begin{example}\label{ex:q}
Let $V=\C^2$, $e=(1,0)^t$, $v=(3,4)^t/5$ and set
\[
 B_0=\tfrac14 ee^*,\qquad B_1=\tfrac14 vv^*,\qquad
 B_2=\tfrac1{10}I.
\]
The two rank-one inputs do not commute: the $(1,2)$ entry of their
commutator is $3/100$. The sum of their ranks, including $B_2$, is
four. Since $ee^*+vv^*$ has eigenvalues $1\pm3/5$, the eigenvalues
of $T$ are $1/5$ and $1/2$. Equation~\eqref{eq:sig} gives the
following signatures on the quotient.
\[
\begin{array}{c|ccccc}
 c&(0,1/5)&1/5&(1/5,1/2)&1/2&(1/2,\infty)\\ \hline
 (n_+,n_-)&(2,2)&(2,1)&(3,1)&(3,0)&(4,0)
\end{array}
\]
Take $q=1/2$ and $\lambda_q=1$. The value $c=1/2$ gives a
three-dimensional positive definite middle space. Both local kernels
and the diagonal kernel contribute to this quotient of $\C^6$.
\end{example}

\subsection{A two-qubit circuit for a local residue}
The following example illustrates how a local residue on the middle
quotient can be read out when its monodromy gate is supplied as an input.
\begin{example}\label{ex:local-residue-circuit}
In a rank-one KZ system take $m=2^n$, $B_i=A_{0i}=a>0$,
$A_{ij}=0$ for $1\le i<j\le m$, and $\mu=-c=-ma$.
Label the coordinates of $W=\C^m$ in binary order. By \eqref{eq:blocks},
\[
 H_c=a(I-|s\rangle\langle s|),\qquad
 |s\rangle=m^{-1/2}\sum_{x\in\{0,1\}^n}|x\rangle .
\]
Thus $[w]\mapsto\sqrt a\,(I-|s\rangle\langle s|)w$ identifies the
middle quotient isometrically with $s^\perp$. The Hadamard gate satisfies
$\mathsf H^{\otimes n}|0^n\rangle=|s\rangle$, so the states
$\mathsf H^{\otimes n}|k\rangle$ for $k\ne 0^n$ form an orthonormal
basis of $s^\perp$. Each takes one layer of $n$ Hadamard gates to prepare
from its label $|k\rangle$.

For $m=4$, the quotient occupies three dimensions of a two-qubit register.
With $|\pm\rangle=(|0\rangle\pm|1\rangle)/\sqrt2$, the transverse residue
\eqref{eq:trans} is
\[
 R_{12}=a(|00\rangle-|01\rangle)(\langle00|-\langle01|)
       =2a\,|0-\rangle\langle0-| .
\]
For the local Euler equation $dY/dz=(R_{12}/z)Y$, a counterclockwise turn
gives $U_a=\exp(2\pi iR_{12})$. Given this gate with $0<a<1/4$ unknown,
apply it once per shot.
\begin{center}
\begin{tikzpicture}[x=1cm,y=1cm,
  gate/.style={draw,fill=white,minimum height=5mm,minimum width=7mm,inner sep=1pt}]
  \draw (0,0)--(4.85,0);
  \draw (0,-0.70)--(4.85,-0.70);
  \node[anchor=east] at (0,0) {$|0\rangle$};
  \node[anchor=east] at (0,-0.70) {$|1\rangle$};
  \node[gate] at (0.65,0) {$\mathsf H$};
  \node[gate] at (0.65,-0.70) {$\mathsf H$};
  \node[gate,minimum height=12mm,minimum width=10mm] at (2,-0.35) {$U_a$};
  \node[gate] at (3.18,0) {$\mathsf H$};
  \node[gate,minimum width=10mm] at (4.55,0) {$M_Z$};
\end{tikzpicture}
\end{center}
The loop sends the prepared state $|+\rangle|-\rangle\in s^\perp$ to
$(e^{4\pi ia}|0\rangle+|1\rangle)|-\rangle/\sqrt2$. After the final Hadamard,
the first qubit gives outcome $1$ with probability
\[
 p_1=\left|\frac{e^{4\pi ia}-1}{2}\right|^2=\sin^2(2\pi a),
 \qquad a=\frac{\arcsin\sqrt{p_1}}{2\pi}.
\]
Use the observed frequency for $p_1$ to estimate $a$.
To test a known $a$, conjugate the controlled-phase gate $\diag(1,1,1,e^{4\pi ia})$
by $X\otimes\mathsf H$. For instance, $a=1/12$ gives $p_1=1/4$.
\end{example}

\section{The quadratic term of the multiplicative KLM form}
\label{sec:tangent}
Exponentiating the self-adjoint input coefficients gives a unitary
tuple for which the additive form appears in the Taylor expansion of
the multiplicative KLM form.

\begin{definition}[Ambient KLM form]\label{def:klmform}
Let $g_j$ be unitary, $|\lambda|=1$, $\lambda\ne1$, and choose a square
root $s$ of $\lambda$. In the convention of \cite[eq.~(3)]{Neg},
the ambient KLM Hermitian matrix $\widetilde H$ has blocks
\[
   \widetilde H_{jk}=\epsilon_{jk}(g_j^{-1}-I)(g_k-I), \qquad
   \epsilon_{jk}=
     \begin{cases}s^{-1},&j<k,\\ s,&j>k,\end{cases}
\]
for $j\ne k$, and $\widetilde H_{jj}=s^{-1}(g_j^{-1}-\lambda I)(g_j-I)$.
\end{definition}
Changing the sign of $s$ changes the sign of the whole matrix.

\begin{theorem}\label{th:tangent}
Let $B_j=B_j^*$ and $c\in\R^\times$. Set
$g_j(t)=\exp(itB_j)$, $\lambda(t)=\exp(-itc)$,
$s(t)=\exp(-itc/2)$. In operator norm on $V^m$ one has
\begin{equation}\label{eq:tangent}
 \widetilde H(t)=t^2(S^*S-cD)+O(t^3)
                =-ct^2 H_c+O(t^3).
\end{equation}
If $c\notin\operatorname{Spec}(T)$, then, for all sufficiently small
nonzero real $t$, the two ambient forms have the same radical $\ker D$,
and $-\widetilde H(t)/(ct^2)$ has the same inertia as $H_c$.
If the $B_j$ commute pairwise, the inertia conclusion also holds at
$c\in\operatorname{Spec}(T)$ for sufficiently small nonzero $t$.
\end{theorem}
\begin{proof}
For $j\ne k$ the product of the linear terms is
$(-iB_j)(iB_k)t^2=t^2B_jB_k$. Both square-root prefactors equal
$1+O(t)$. For $j=k$ the product is
$t^2(B_j-cI)B_j+O(t^3)$. These are the blocks of $S^*S-cD$.
The rightmost factor $g_j-I$ in column $j$ shows that
$E_j\ker B_j$ lies in the radical for every $t$.
For $c\notin\operatorname{Spec}(T)$, $H_c$ restricts to a nondegenerate form on
$(\ker D)^\perp$. Norm convergence on this fixed complement implies
stability of its inertia.

For commuting tuples $\lambda(t)g_1(t)\cdots g_m(t)=\exp(it(T-cI))$.
For small nonzero $t$ its fixed space is $\ker(T-cI)$, and likewise
$\ker(g_j(t)-I)=\ker B_j$. By the multiplicative radical formula
\cite[eq.~(2)]{Neg}, the nullities of the two forms therefore agree.
Norm convergence preserves the signs of the nonzero eigenvalues for
sufficiently small $t$, and equality of the nullities accounts for the
remaining eigenvalues.
\end{proof}

Here $s(t)$ is the local branch with $s(0)=1$. An
eigenangle convention with values in $[0,1)$ can use the opposite
square root on one side of the identity. Formula~\eqref{eq:tangent}
fixes the sign before taking the limit. The expansion also holds for noncommuting $B_j$ and for
$c\in\operatorname{Spec}(T)$. At such a value of $c$, equality of
the two ranks requires a separate check.

There is a parallel first-order identity for the free-group generators.
In the multiplicative convolution row convention \cite{NegMC,DR}, $N_i-I$
has only row $i$ nonzero. Its block $j$ is $\lambda(g_j-I)$ for $j<i$,
$\lambda g_i-I$ for $j=i$, and $g_j-I$ for $j>i$. Consequently
\begin{equation}\label{eq:first}
                   N_i(t)=I+itR_i+O(t^2).
\end{equation}
The first-order coefficient gives the operator, and the quadratic
coefficient \eqref{eq:tangent} gives its form. Indeed the coefficient
of $t^3$ in $N_i(t)^*\widetilde H(t)N_i(t)-\widetilde H(t)$
is $-ic(H_cR_i-R_i^*H_c)$. Thus multiplicative invariance also
recovers the infinitesimal adjoint identity \eqref{eq:row}.

\begin{example}[Variation of the singular parameter for noncommuting residues]\label{ex:shift}
Use Example~\ref{ex:q}, and now vary the exponentiation parameter $t$.
For small $t>0$ the two eigenphases of $g_0(t)g_1(t)g_2(t)$, divided
by $t$, are
\begin{equation}\label{eq:wallshift}
 c_\pm(t)=\frac7{20}\ \pm\ \frac2t
      \arcsin\!\left(\frac35\sin\frac t8\right).
\end{equation}
Indeed, set $u=e^{it/4}$. In the first two factors
$e^{itP/4}=I+(u-1)P$ for $P=ee^*,vv^*$. The trace of their product
is $2+2(u-1)+(9/25)(u-1)^2$ and its determinant is $u^2$.
After division of the product by $u$ the determinant is one and the
half trace is $1-(18/25)\sin^2(t/8)$. Its eigenvalues are
$e^{\pm i\phi}$ where $\phi=2\arcsin((3/5)\sin(t/8))$.
The last scalar factor contributes $e^{it/10}$, giving
\eqref{eq:wallshift}. In particular,
\[
 c_+(t)=\tfrac12-\tfrac{t^2}{4000}+O(t^4),\qquad
 c_-(t)=\tfrac15+\tfrac{t^2}{4000}+O(t^4).
\]
At $c=1/2$ the additive middle space has dimension three. For all sufficiently small $t>0$, neither eigenphase divided by
$t$ equals $1/2$. The multiplicative middle space then has dimension
four. Along $c=c_+(t)$ the product $\lambda(t)g_0(t)g_1(t)g_2(t)$ again has eigenvalue one.
\end{example}

\section{Integral structures and symmetry}\label{sec:integral}
\subsection{The lattice of a radical quotient}
Let $K\subset\C$ be a number field stable under complex conjugation,
with ring of integers $\mathcal O$. Work at the specified embedding.
\begin{definition}[Full lattice and integral middle space]\label{def:middlelattice}
A full lattice in $W_K$ is a finitely generated $\mathcal O$-submodule
$\Lambda\subset W_K$ with $K\Lambda=W_K$. Suppose the spaces, metrics
and maps in \eqref{eq:data} are defined over $K$, and let $N_K$ be the
radical of $H$ over $K$. The integral middle space is
\begin{equation}\label{eq:lattice}
                 \Lambda_M=\Lambda/(\Lambda\cap N_K).
\end{equation}
\end{definition}

Adjoints in arbitrary bases use the input Hermitian matrices.
The extension of $\Lambda_M$ to $\C$ recovers $M$, since kernels of
matrices over $K$ commute with field extension.

\begin{proposition}\label{prop:lattice}
The module \eqref{eq:lattice} is torsion-free and projective over
$\mathcal O$. Every lattice-preserving coefficient $X$ satisfying $HX=X^*H$ induces
an endomorphism of the quotient lattice. Every lattice automorphism
preserving $H$ likewise induces an automorphism of the quotient lattice.
If the form is integral on $\Lambda$, its induced form on $\Lambda_M$
is integral. In general a positive rational integer clears its
denominators and preserves its inertia at the chosen embedding.
\end{proposition}
\begin{proof}
If $0\ne a\in\mathcal O$ and $aw\in N_K$, then $w\in N_K$. Thus $\Lambda\cap N_K$ is saturated in $\Lambda$, so the quotient is
torsion-free. Finitely generated torsion-free modules over a Dedekind
domain are projective. The stated operators preserve the
radical, and the values of the form are independent of representatives.
A finite set of lattice generators gives finitely many pairings in $K$;
some positive integer sends all of these into $\mathcal O$.
\end{proof}

In particular, the entries of $C^{-1}$ need not lie in $\mathcal O$. Integrality of any proposed group generators is a
separate requirement. For instance, if $X=X^{\dagger}$ for a positive
quotient metric, its Cayley transform $(I-iX)(I+iX)^{-1}$ is an
isometry: both factors commute and are adjoints, and $I+iX$ is
invertible because the spectrum is real. A tuple of such Cayley transforms therefore defines a unitary
representation of the free group.
A prescribed relation descends exactly when the corresponding word
acts as the identity on the quotient. For lattice automorphisms
$g_j$ already acting on $\Lambda$ and preserving $H$, the same test
has the integral form
\begin{equation}\label{eq:rel}
 \bigl(r(g)-I\bigr)\Lambda\ \subseteq\ \Lambda\cap N_K .
\end{equation}
Imposing a list of Artin or other relators therefore reduces to testing
\eqref{eq:rel} on generators of the lattice. For a group permuting
parameters through $\pi:G\to\mathfrak S_m$, a fixed parameter tuple has
stabilizer $\pi^{-1}(\operatorname{Stab}(\mathbf b))$. The subgroup
$\ker\pi$ fixes every ordered tuple. Extension beyond the stabilizer
requires maps between the corresponding fibers.

\begin{example}[The root lattice $A_m$]\label{ex:root}
Take $V=\C$, $B_i=1$ and $c=-1$. On $\Z^m$ the Hermitian matrix is
$I+\mathbf1\mathbf1^t$. Let $s_i=I-R_i$ with $R_i$ as in
\eqref{eq:row}. The homomorphism
\[
 (w_1,\ldots,w_m)\longmapsto
       (-\textstyle\sum_iw_i,w_1,\ldots,w_m)
\]
is an integral isometry from the middle lattice to the sum-zero
sublattice of $\Z^{m+1}$ with its Euclidean form. Under this map $s_i$
interchanges coordinates $0$ and $i$. Hence the $s_i$ generate
$\mathfrak S_{m+1}$ on the positive middle lattice. The usual adjacent
transpositions are $s_1$ and $s_i s_{i+1}s_i$ for $1\le i<m$,
so the Coxeter and braid relations hold integrally. This example
constructs group isometries directly from the coefficient rows.
\end{example}

\subsection{Integral Pochhammer cycles for a presented group}
We now start with a multiplicative representation. The question here
is whether the cycles that generate a middle space over a field also
generate its natural integral lattice. Retain the ring $\mathcal O$ and
let $\Gamma=\langle x_1,\ldots,x_n\mid r_1,\ldots,r_b\rangle$ act by
automorphisms $g_i$ on a finite free $\mathcal O$-module $A$.
Set $\Delta=\Gamma*\langle z\rangle$ and extend the action by
$z\mapsto\lambda I$, where $\lambda\in\mathcal O^\times$ and
$\delta=\lambda-1\ne0$. Write $\bar A=A/\delta A$.
On $\bar A$ the action factors through the retraction
$p:\Delta\to\Gamma$ killing $z$. There is a canonical map
\[
 p_*:H_1(\Delta;A)\longrightarrow H_1(\Gamma;\bar A)
\]
induced by $p$ and coefficient reduction.
\begin{definition}[Pochhammer cycles and the peripheral quotient]\label{def:peripheral}
For each $i$ the commutator $[z,x_i]=zx_i z^{-1}x_i^{-1}$ acts
trivially on $A$. It therefore gives a cycle for each coefficient in
$A$. Let $P\subset H_1(\Delta;A)$ be the $\mathcal O$-submodule generated
by these Pochhammer cycles. Hiroe and Negami
\cite[Def.~2.13]{HN} use $z^{-1}x_i^{-1}zx_i$. Conjugating their
word by $zx_i$ gives ours. This transports the coefficient by an
automorphism of $A$, so both choices give the same span.
For a finite collection $\mathcal P$ of
peripheral subgroup maps into $\Delta$, let $B_{\mathcal P}$ be the sum of the images of the induced maps on first homology,
using the restricted coefficients on each peripheral subgroup. Put $B_{\mathcal P}^{\mathrm{sat}}=\{x\in H_1(\Delta;A): ax\in B_{\mathcal P}
\text{ for some }0\ne a\in\mathcal O\}$. The saturated peripheral
middle lattice is $M_{\mathcal P}=H_1(\Delta;A)/B_{\mathcal P}^{\mathrm{sat}}$.
\end{definition}

\begin{theorem}\label{th:integral}
The map $p_*$ induces a natural isomorphism
\begin{equation}\label{eq:defect}
 H_1(\Delta;A)/P\simeq H_1(\Gamma;\bar A).
\end{equation}
In particular, $P$ is independent of the chosen generating set of
$\Gamma$ for the specified splitting. For the prescribed peripheral maps,
\begin{equation}\label{eq:perdef}
 M_{\mathcal P}/\im(P)\simeq H_1(\Gamma;\bar A)/p_*(B_{\mathcal P}^{\mathrm{sat}}).
\end{equation}
The quotient $M_{\mathcal P}/\im(P)$ is killed by $\delta$, and is supported only at primes
dividing $(\delta)$.
\end{theorem}
\begin{proof}
Use the presentation complex, with the $z$ edge first. Its degree-one
boundary for $\Delta$ is
\[
  (a,x)\longmapsto\delta a+D_1 x,\qquad
  D_1(x)=\sum_{i=1}^n (g_i-I)x_i.
\]
The boundaries from degree two have form $(0,D_2 y)$, with $D_2$ the
relator boundary for $\Gamma$. This computes group homology in degree
one: cells of dimension at least three can be attached to construct a
classifying space. As $\delta$ acts injectively on $A$, projection to
$x$ identifies the cycles with the vectors $x\in A^n$ satisfying
$D_1x\in\delta A$. Reduction of $x$ is exactly the map $p_*$.
Every cycle modulo $\delta$ lifts to such an $x$. A class is in the
kernel if its tail reduces to a relator boundary. Lifting that boundary
shows that the class has a representative with $x\in\delta A^n$.

We use the product rule of the free differential calculus
\cite[\S\S1--2]{Fox} in the convention that traverses the right factor
first. Thus $[uv]a=[v]a+[u]\rho(v)a$ for chains. The tail of $[z,x_i]$ with coefficient
$a$ has only its $i$th component nonzero, equal to
$(\lambda^{-1}-1)g_i^{-1}a$. These vectors fill $\delta A$ in that
component. They generate all classes in $\ker p_*$ because the first
coordinate of a cycle is determined by its tail. This proves
\eqref{eq:defect}; naturality follows from that of $p_*$ for maps
compatible with retraction and coefficient reduction. Dividing further
by $B_{\mathcal P}^{\mathrm{sat}}$ gives \eqref{eq:perdef}. The right side has
coefficients annihilated by $\delta$.
\end{proof}

For a compact oriented surface $\Sigma$ with boundary and a unitary
local system $A_K$, take the boundary circles as peripherals. The exact
sequence of the pair identifies the middle space over $K$ with
\[
 \im\!\left(H_1(\Sigma;A_K)\longrightarrow
                    H_1(\Sigma,\partial\Sigma;A_K)\right).
\]
Indeed the kernel is the boundary image. For the dual description in
parabolic cohomology of a punctured sphere see \cite[\S2]{DW}.
Poincar\'e--Lefschetz
duality identifies the absolute and relative spaces as Hermitian duals
using the coefficient metric. Pulling their intersection pairing back
to absolute homology gives a skew-Hermitian pairing with this same
kernel. At the complex embedding multiplication by $i$ makes it
Hermitian. It is invariant under orientation-preserving symmetries
lifting to isometries of the local system. By localization and exactness, the map from integral homology to the
middle space over $K$ has kernel $B_{\mathcal P}^{\mathrm{sat}}$. Its image is
therefore the lattice in \eqref{eq:perdef}, independently of the
choice of Pochhammer generators. For a general presented group the peripheral
quotient still exists. A duality pairing is extra data there.

In the case of the punctured plane take $\Gamma=F_n$, let $z$ encircle
the additional moving puncture and include the peripheral circle at
infinity. Relative homology of a compact core identifies with
Borel--Moore homology of the interior. With the Kummer-twisted local
system and braid action of \cite[Thm.~0.4]{HN}, the middle image over
$K$ is the KLM representation. Equation~\eqref{eq:perdef} now measures
the difference between the commutator-generated lattice and the full
middle image lattice under this homological realization. A change of
normalization of the cycles can change their integral span.

\begin{example}[Pochhammer cycles generate a sublattice of index $25$]\label{ex:25}
Take $\mathcal O=\Z[\zeta_5]$, $A=\mathcal O^2$, $\Gamma=F_2$,
$\lambda=\zeta_5$ and
\[
 g_1=\diag(-1,\zeta_5),\qquad g_2=\diag(\zeta_5,-1).
\]
These matrices are unitary and commute. The coordinate swap
intertwines the interchange of the two local labels. At the three
finite points the local matrices $g_1,g_2,\lambda I$ have no fixed
vectors, and the product $\lambda g_1g_2=-\zeta_5^2 I$ gives the
same conclusion at infinity. Hence every peripheral $H_1$ is zero.
Here $H_1(\Delta;A)$ is torsion-free since there are no relator boundaries, so
$B_{\mathcal P}^{\mathrm{sat}}=0$. The reduction $\mathcal O/(\zeta_5-1)$ is
$\mathbb F_5$. In the reduced boundary map
$\mathbb F_5^4\to\mathbb F_5^2$, the two nonzero rows have disjoint
support and respective nonzero coefficients $-2,-2$. Its kernel is
$\mathbb F_5^2$. By \eqref{eq:perdef} the Pochhammer lattice has
index $25$ in the middle lattice, although the two have identical
scalar extensions to $\Q(\zeta_5)$.

One can now impose the length-two Artin relation $x_1x_2=x_2x_1$,
since these same matrices commute. This replaces $F_2$ by $\Z^2$.
The torus supplies a degree-two boundary with components
$(I-\bar g_2,\bar g_1-I)$, up to the orientation of the two-cell.
Its image spans the two unused coordinate directions in the preceding
kernel, since $2$ is a unit in $\mathbb F_5$. Thus
$H_1(\Z^2;\bar A)=0$. Hence $H_1(\Delta;A)/P=0$ for $\Delta=\Z^2*\langle z\rangle$, even before taking the peripheral quotient.
\end{example}

\section{KLM representations of marked quotients}\label{sec:quot}
Fix a number field $K\subset\C$ stable under complex conjugation,
its ring of integers $\mathcal O$, a unit $\lambda\in\mathcal O^\times$
of absolute value one at the specified embedding, and a square root
$s=\lambda^{1/2}\in K$. Assume $\lambda\ne1$.

\begin{definition}[Marked input]\label{def:marked}
Let $F_n$ have ordered free generators $x_1,\ldots,x_n$. Specify a
list of relators $\mathcal R$ with normal closure $N$, and put
$\Gamma=F_n/N$. The marking of $\Gamma$ is the ordered tuple of
images of the $x_i$. We use $(x_1\cdots x_n)^{-1}$ as the peripheral
word at infinity. Take $B=1$ to work with the marked group $\Gamma$
alone. More generally, let $B$ be a group equipped with a homomorphism
$\beta:B\to B_n$ such that the induced Artin action preserves $N$.
This is the setting of \cite[Def.~1.1]{HN} and includes inclusions of
braid subgroups as well as homomorphisms with kernel.

Let $V$ be a $K$-representation of $\Gamma\rtimes B$ preserving a
$K$-valued Hermitian form that is positive definite at the chosen
embedding, and let
$A\subset V$ be a stable full $\mathcal O$-lattice.
\end{definition}
Inflate the representation to $F_n\rtimes B$. In the multiplicative KLM construction
\cite{HN,NegMC,Neg}, put
\[
     M=\KLM_\lambda(V),\qquad
     h=\text{the induced KLM form},\qquad
     L=\im(A^n\longrightarrow M).
\]
Here $h$ is nondegenerate and can be indefinite. In the block
formula of Section~\ref{sec:tangent} use the input metric in place
of the orthonormal identification with the dual. The KLM row matrices
and the braid action preserve the image lattice: the free rows and
their inverses use $g_i^{\pm1},\lambda^{\pm1}$, and the
action of $b\in B$ sends $f\otimes v$ in the augmentation-ideal
construction to $\beta(b)(f)\otimes bv$. In particular,
$\ker\beta$ acts on the coefficient factor through its given action
on $A$. The radical is
defined over $K$. Denote the full action on $M$ by $\rho$.

\begin{definition}[Middle quotient for prescribed relations]\label{def:markedquotient}
Put $J=\sum_{u\in N}\im(\rho(u)-I)$ and $M_N=M/J$ over $K$.
Define the output space and its lattice by
\[
 Q=\im(M^N\longrightarrow M_N),\qquad
 L_Q=Q\cap\im(L\longrightarrow M_N).
\]
\end{definition}

\begin{theorem}[Descent to a marked quotient]\label{th:descent}
The group $\Gamma\rtimes B$ acts on $Q$ and preserves the full
torsion-free lattice $L_Q$. For $v,w\in M^N$, the formula
$h_Q([v],[w])=h(v,w)$ defines a nondegenerate invariant Hermitian
form on $Q$. Write $(p,q)$ for the signature of $h$
and $(a,b,t)$ for the inertia of $h|_J$. Then
\begin{equation}\label{eq:quotindices}
             \operatorname{sig}(h_Q)=(p-a-t,\ q-b-t).
\end{equation}
In particular the induced form is positive definite precisely when
$q=b+t$ (and has positive dimension when $p>a+t$). A positive
integer multiple is integral on $L_Q$.

For a fixed marking, lattice-preserving intertwiners induce maps of
the output lattices functorially, and input isometries induce output
isometries. The construction depends on the relators only through
their normal closure. With
$N=1$ the identities $Q=M$, $L_Q=L$, $h_Q=h$ recover the free
generators and the full prescribed braid action.
\end{theorem}
\begin{proof}
For every $u\in N$, the orthogonal complement of
$\im(\rho(u)-I)$ is $\ker(\rho(u^{-1})-I)$. Hence $M^N=J^\perp$.
Normality in $F_n\rtimes B$ makes the fixed space and $J$ stable.
Thus $Q=J^\perp/(J\cap J^\perp)$ and the representative formula is
exactly the nondegenerate quotient form. The subgroup $N$ acts
trivially there, so the action factors through $\Gamma\rtimes B$.
The image of $L$ in $M_N$ is a full lattice, and intersection with
the $K$-subspace $Q$ gives a full lattice there. Denominators of
finitely many pairings can be cleared as in Proposition~\ref{prop:lattice}.

Split off from $J$ a nondegenerate summand of signature $(a,b)$.
In its orthogonal complement the remaining $t$ dimensions of $J$
are totally isotropic. Completing a basis of this subspace to
$t$ hyperbolic pairs splits off signature $(t,t)$: choose dual
vectors by nondegeneracy, then use half the entries of their Hermitian matrix of pairings
to adjust them in the isotropic directions and make the dual vectors isotropic. The
remaining nondegenerate space is isometric to $Q$, proving
\eqref{eq:quotindices}. An intertwiner of input representations induces the usual
KLM intertwiner, preserving fixed spaces and sending $J$ into $J'$.
It preserves the specified image lattices and hence their intersections.
This proves functoriality and the specialization statements.
\end{proof}

For a finite presentation $J$ is computable without enumerating $N$.
Start with the sum of the images of $\rho(r)-I$ for
$r\in\mathcal R$, and close this subspace under the $n$ meridian
operators and their inverses until its dimension stabilizes. The resulting subspace is the smallest $F_n$-stable subspace on whose
quotient all the relators act trivially. It is therefore $J$. At most $\dim M$ strict dimension increases are
possible. When $h>0$, $M=J\perp J^\perp$ and $L_Q$ corresponds in
$J^\perp$ to the orthogonal projection of $L$. Intersection with
$J^\perp$ may give a smaller lattice. The map from invariants to coinvariants and the inertia calculation
apply to any representation preserving a nondegenerate Hermitian form.
For KLM we can compute $J$ directly from the input.
After extension to $\C$, the unitary input splits as an orthogonal sum
of irreducible $F_n$-representations. Call a character \emph{almost trivial}
if at most one marked generator acts nontrivially. If an irreducible
summand $Y$ is not such a character, every intertwiner between $Y$ and
an almost trivial character, in either direction, vanishes because a nonzero
intertwiner would be an isomorphism. Thus $Y$ belongs to
$\mathrm{Mod}^{\rm NT}$ and its middle representation is irreducible
\cite[Defs.~5.1--5.2, Cor.~5.15]{HN}. If $Y$ is an almost trivial
character, the local kernels kill all but at most one coordinate,
so its middle space has dimension zero or one. This proves
irreducibility of every nonzero middle summand, including the exceptional cases.
On each such summand the span of the $N$-differences is zero or the
whole summand. Thus $t=0$ for these representations even for general relators.
For cyclic orders the question is which summands survive. Over the
coefficient field, the criterion for nonexceptional irreducible summands
also follows from the local Jordan-block rule \cite[Lem.~2.3]{DRJ} together
with irreducibility. Our proof uses the convolution rows and also accounts for the
exceptional characters, the braid action, and the marked lattice.

\subsection{An ideal in the group ring of the input}
There is a single test on the input for all the relations at once.
Put $R_\Gamma=\mathcal O[\Gamma]$ with involution extending conjugation by
$\gamma^*=\gamma^{-1}$. Write $u_i$ for the image of $x_i$ in
$\Gamma$, also viewed as a unit of $R_\Gamma$, and extend $*$ to matrices by applying the involution entrywise and
transposing.

\begin{definition}[Matrices over the group ring and the input ideal]\label{def:inputideal}
Define $\mathcal T:F_n\to\mathrm{GL}_n(R_\Gamma)$ by
$\mathcal T(x_i)=I+E_i q_i$ where $E_i$ is the $i$th coordinate
column, and the entries of row $q_i$ are
\[
 (q_i)_j=
 \begin{cases}
 \lambda(u_j-1)&j<i,\\
 \lambda u_i-1&j=i,\\
 u_j-1&j>i .
 \end{cases}
\]
Let $\mathcal H=\mathcal H^*$
be the block formula of Section~\ref{sec:tangent} with $g_i=u_i$.
Thus $\mathcal H_{ii}=s^{-1}(u_i^{-1}-\lambda)(u_i-1)$ and
$\mathcal H_{ij}=s^{-1}(u_i^{-1}-1)(u_j-1)$ for $i<j$.
For $w\in F_n$ set $\mathcal D(w)=\mathcal H(\mathcal T(w)-I)$.
The input ideal and its annihilator in $V$ are
\begin{equation}\label{eq:ideal}
 \mathfrak a_{\lambda,N}=
 \sum_{\substack{r\in\mathcal R\\1\le i,j\le n}}
             R_\Gamma\,\mathcal D(r)_{ij}\,R_\Gamma
       \ \subset R_\Gamma,\qquad
 V_N=\{v\in V:\mathfrak a_{\lambda,N}v=0\}.
\end{equation}
\end{definition}
The matrices $\mathcal T(x_i)$ are invertible over $R_\Gamma$ since their
diagonal entry in the modified row is a unit. The entries lie in the group ring of $\Gamma$, while the argument
of $\mathcal T$ is a word in $F_n$.

\begin{theorem}[The input ideal and the output lattice]\label{th:ideal}
The ideal in \eqref{eq:ideal} is independent of the chosen list of normal generators of
$N$ for the given marking and parameter. For an input representation with a positive definite invariant form
and a lattice as above, $V_N$ is stable under $\Gamma\rtimes B$.
Let $\pi_N:V\to V_N$ be orthogonal projection and let $h_{\KLM,V_N}$
be the KLM form induced by the restriction of the input form to $V_N$. Then
\begin{equation}\label{eq:idealthree}
 (Q,h_Q,L_Q)\simeq
 \bigl(\KLM_\lambda(V_N),\
       h_{\KLM,V_N},\
       \im((\pi_N A)^n\to\KLM_\lambda(V_N))\bigr)
\end{equation}
equivariantly for $\Gamma\rtimes B$. In particular, the induced generators are exactly the operators
obtained by applying KLM to $V_N$. For
$\lambda=e^{2\pi i \ell}$, $0<\ell<1$, take $s=e^{\pi i \ell}$. Write $g_i$
for the input meridians restricted to $V_N$, $P=\lambda g_1\cdots g_n$,
$R_0=\sum_i\rk(g_i-I)$ and $r=\dim\ker(P-I)$. If $\theta(U)$ is the
sum, with multiplicities, of the eigenangles of $U$ in $[0,1)$, then
\[
 \kappa=\ell\dim V_N+\sum_i\theta(g_i)-\theta(P),\qquad
 \operatorname{sig}(h_Q)=(R_0-\kappa,\kappa-r).
\]
The output is nonzero and positive definite exactly when $R_0>r$
and $\kappa=r$. Changing the square root negates the form.
When $N=1$ the ideal is zero and \eqref{eq:idealthree} recovers the
whole KLM representation before imposing relations, its form and its marked lattice.
\end{theorem}
\begin{proof}
The identities $\mathcal H E_i=q_i^*s(u_i-1)$ and
\[
 s(u_i-1)+s^{-1}(u_i^{-1}-1)+\mathcal H_{ii}=0
\]
give $\mathcal T(x_i)^*\mathcal H\mathcal T(x_i)=\mathcal H$ by
expansion. These identities hold in the group ring before evaluation
on an input representation. It follows that
\begin{align*}
 \mathcal D(uv)&=\mathcal D(u)\mathcal T(v)+\mathcal D(v),\\
 \mathcal D(u^{-1})&=-\mathcal D(u)\mathcal T(u)^{-1},\\
 \mathcal D(uru^{-1})&=(\mathcal T(u)^{-1})^*
                               \mathcal D(r)\mathcal T(u)^{-1}.
\end{align*}
Thus the same two-sided ideal is generated if $r$ ranges over all of
$N$, proving independence from the relator list.

Evaluating $\mathcal H$ on $V^n$ gives the operator of the KLM form
relative to the input metric. Its kernel is the sum of the local and
infinity kernels by the radical formula \cite[Thm.~9]{NegMC},
also recorded in \cite[eq.~(2)]{Neg}. For an $F_n$-submodule $Y$,
positivity of the input metric gives $V=Y\perp Y^\perp$ over $K$.
Each block of the evaluated $\mathcal H$ preserves both summands,
since it is a linear combination of words in the input operators.
Thus the ambient form and its radical split, and the two ordinary
middle spaces embed in $M$ as $h$-orthogonal nondegenerate summands,
including when $h$ is indefinite. A word
$r$ acts trivially on $\KLM_\lambda(Y)$ exactly when the image of
$(\mathcal T(r)-I)Y^n$ lies in that middle kernel, equivalently when
each $\mathcal D(r)_{ij}$ kills $Y$. Since $Y$ is an $R_\Gamma$-submodule,
this is the condition $\mathfrak a_{\lambda,N}Y=0$ when imposed on
all relators. Thus $V_N$ is the largest $F_n$-submodule on which
the middle action respects the relators. The KLM action of $b\in B$
transports the middle subspace for $Y$ onto that for $bY$
(by the augmentation-ideal construction \cite{HN}); $b$ preserves
$N$. This gives $B$-stability of $V_N$.

Extend scalars along $K\hookrightarrow\C$. Taking the common kernel that defines $V_N$ commutes with scalar
extension. Indeed, the evaluated ideal spans a finite-dimensional
$K$-subspace of $\operatorname{End}_K(V)$, so it suffices to
intersect the kernels of a basis of this subspace. The same argument
applies to the spans defining $J$, and the middle kernel commutes
with extension as the kernel of the evaluated matrix.
Decompose $(V_N^\perp)_{\C}$ into orthogonal irreducible complex
$F_n$-modules $Y$. On each such $Y$ some relator acts
nontrivially on $M(Y)=\KLM_\lambda(Y)$, by the maximality just
proved. Since $N$ is normal, the span of the $N$-differences inside
$M(Y)$ is a nonzero $F_n$-submodule. By the two cases above, $M(Y)$
is irreducible, so this span is all of $M(Y)$. On $\KLM_\lambda(V_N)$ every such difference
vanishes. Additivity gives
$J_{\C}=\KLM_\lambda(V_N^\perp)_{\C}$ inside $M_{\C}$.
Both are extensions of $K$-subspaces, so the equality holds over
$K$. Coordinatewise
projection on $V^n$ commutes with passage to this middle splitting,
so the image of $A^n$ in the quotient is exactly the image of
$(\pi_N A)^n$. This proves \eqref{eq:idealthree}. The signature assertion
is \cite[Thm.~1.1]{Neg}.
\end{proof}

\begin{definition}[Projected and intersection lattices]\label{def:projection}
For a $K$-subrepresentation $Y\subset V$ with orthogonal projection
$\pi_Y$, the projected and intersection lattices in $Y$ are $\pi_Y A$
and $A\cap Y$, respectively. Their images under the middle quotient
$q_Y:Y^n\to\KLM_\lambda(Y)$ are
$L_{\rm proj}=q_Y((\pi_Y A)^n)$ and $L_{\rm int}=q_Y((A\cap Y)^n)$.
Two Hermitian lattices are similar if a complex-linear isomorphism
carries one onto the other and multiplies their forms by a positive
real scalar.
\end{definition}

For a finite relator list, intersect the kernels in $V$ of the
finitely many evaluated entries $\mathcal D(r)_{ij}$. To obtain the largest $F_n$-stable subspace of this intersection,
successively intersect it with its translates under the generators. This computes $V_N$ entirely inside $V$. The annihilator may include
summands with zero KLM representation. Omitting these summands in the
input test leaves the projected middle lattice unchanged.

Adding relators gives a filtration for any fixed representation on which the
new relators already hold. Indeed if $N\subseteq N'$ are preserved by
$B$, the ring map $\mathcal O[F_n/N]\to\mathcal O[F_n/N']$ sends
$\mathfrak a_{\lambda,N}$ into $\mathfrak a_{\lambda,N'}$: the
matrices $\mathcal T(x_i)$ specialize and every old relator lies in $N'$.
Consequently $V_{N'}\subseteq V_N$. Orthogonal projection onto $V_{N'}$
factors through projection onto $V_N$. It therefore induces a
surjection of the marked output lattices $L_{Q_N}\to L_{Q_{N'}}$,
equivariant for $(F_n/N)\rtimes B$ by inflation. These maps compose
for chains of added relations.

\subsection{Meridians of prescribed order}
\begin{definition}[The input subspace for prescribed meridian orders]\label{def:orders}
Let $d_i$ be an integer at least two, or $\infty$, and set
\[
 \Gamma_{\mathbf d}=
 \langle x_1,\ldots,x_n\mid x_i^{d_i}=1\ (d_i<\infty)\rangle .
\]
Write $C_d$ for the cyclic group of order $d$. Each finite $d_i$
contributes a factor $C_{d_i}$ to the free product $\Gamma_{\mathbf d}$,
and each $d_i=\infty$ contributes a factor $\Z$. Take $B$ preserving the orders under
its permutation of the markings. With $g_i$ the input meridians, put
\[
 \mathcal I_{\ne 1}=\{i:d_i<\infty,\ \lambda^{d_i}\ne1\},\qquad
 \mathcal I_1=\{i:d_i<\infty,\ \lambda^{d_i}=1\}.
\]
For a subspace $Y\subset V$, write $K[F_n]Y$ for the submodule
it generates. Set
\begin{equation}\label{eq:good}
 D_* = K[F_n]\!\left(\sum_{i\in\mathcal I_{\ne 1}}(g_i-I)V+
                         \sum_{i\in\mathcal I_1}\ker(\lambda g_i-I)\right),
 \qquad V_* = D_*^\perp .
\end{equation}
\end{definition}

Equivalently, $V_*$ is the largest $F_n$-subrepresentation on which
every $g_i$ with $i\in\mathcal I_{\ne 1}$ acts trivially and no $g_i$ with $i\in\mathcal I_1$
has the eigenvalue $\lambda^{-1}$. This follows from unitarity and
the orthogonal spectral decomposition of each $g_i$. These conditions
are invariant under conjugation of meridians and permutations
preserving their orders. Thus $V_*$ and $D_*$ are $B$-stable. Orthogonal
projection $\pi_*:V\to V_*$ is defined over $K$ and computable by
linear equations.

\begin{theorem}[Cyclic peripheral relations]\label{th:cyclic}
Suppose the input metric is positive definite. For any signature of
the initial KLM form, we have a canonical isometry of Hermitian spaces
with their indicated lattices
\begin{equation}\label{eq:cyclic}
 (Q,h_Q,L_Q)\simeq
 \bigl(M_*,h_{\KLM,V_*},\im((\pi_* A)^n\to M_*)\bigr),
 \qquad M_*=\KLM_\lambda(V_*).
\end{equation}
On the right, the meridians and braids act through the KLM
representation associated with $V_*$. This action factors through
$\Gamma_{\mathbf d}\rtimes B$. In particular
$J=\KLM_\lambda(D_*)$ is a nondegenerate subspace of $M=\KLM_\lambda(V)$.
For the branch $s=e^{\pi i \ell}$, where $\lambda=e^{2\pi i \ell}$ and
$0<\ell<1$, the signature is
\[
 (R_0-\kappa,\kappa-r),\qquad \kappa=\ell\dim V_*+\sum_i\theta(g_i|_{V_*})-\theta(P),
\]
where $P=\lambda(g_1\cdots g_n)|_{V_*}$,
$R_0=\sum_i\rk((g_i-I)|_{V_*})$, $r=\dim\ker(P-I)$, and $\theta$
is the sum of eigenangles in $[0,1)$, with multiplicities. Thus the
form is positive definite on a nonzero output precisely when $\kappa=r$
and $R_0>r$. The other square-root branch negates the form.
\end{theorem}
\begin{proof}
Write $\iota_k:V\to M$ for coordinate inclusion followed by the middle
quotient, and $T_j=\rho(x_j)$. The multiplicative rows give
\begin{equation}\label{eq:propagate}
 T_j\iota_j v=\lambda\iota_j g_jv,\qquad
 (T_j-I)\iota_k v=\alpha_{jk}\iota_j(g_k-I)v \quad(k\ne j),
\end{equation}
where $\alpha_{jk}=\lambda$ for $k<j$ and $1$ for $k>j$. On the
ambient block space write $T_i-I=E_i Q_i$. For $d=d_i<\infty$,
\begin{equation}\label{eq:power}
 T_i^d-I=E_i\left(\sum_{k=0}^{d-1}(\lambda g_i)^k\right)Q_i .
\end{equation}
If $i\in\mathcal I_{\ne 1}$, then $g_i$ acts as the identity on $V_*$. The entire
$i$th coordinate summand therefore lies in the local kernel, so
$T_i=I$ on the middle quotient. If $i\in\mathcal I_1$ the
sum in \eqref{eq:power} vanishes since $(\lambda g_i)^d=I$ and
$\lambda g_i-I$ is invertible on $V_*$. All relators therefore act
trivially on $M_*$.

Put $D'=\{v:\iota_jv\in J\text{ for all }j\}$. It is $F_n$-stable:
for $v\in D'$ the diagonal equation of \eqref{eq:propagate} gives
$\iota_k g_kv\in J$ and the others give
$\iota_j(g_k-I)v\in J$ for $j\ne k$. Finite dimension gives
stability under inverses. Furthermore, if $\iota_i u\in J$, then
$(g_i-I)u\in D'$, by the same equations applied to the $i$th summand.

Decompose the unitary input representation orthogonally over $K$ into irreducible
$F_n$-modules $Y$. Orthogonal complements exist over $K$ by positivity
of the input metric. The KLM form and free meridian action split over
this decomposition, which need not split the braid action or the
lattice. On each $Y$ the contribution to $D_*$ in \eqref{eq:good}
is either zero or all of $Y$. If $i\in\mathcal I_{\ne 1}$ and $g_i|_Y\ne I$, then
$(T_i^{d_i}-I)\iota_i u=(\lambda^{d_i}-1)\iota_i u$ for every
$u\in Y$. Thus $D'$ contains $(g_i-I)Y\ne0$ and hence all of $Y$.

Suppose next that $i\in\mathcal I_1$ and $E=\ker((\lambda g_i-I)|_Y)\ne0$.
The geometric sum in \eqref{eq:power} is $d$ times the orthogonal
projection $P_E$ on $Y$. If $P_E(g_k-I)Y\ne0$ for some $k\ne i$,
choose a nonzero $u$ in that image. Applying \eqref{eq:power} to the $k$th summand shows $\iota_i u\in J$. Then $(g_i-I)u=(\lambda^{-1}-1)u$
is a nonzero element of $D'\cap Y$, so $Y\subseteq D'$.
Otherwise unitarity gives $E\subseteq\ker(g_k-I)$ for all $k\ne i$.
Then $E$ is a nonzero $F_n$-submodule and equals $Y$. On this
exceptional character $g_i=\lambda^{-1}$ and the other $g_k=1$.
Its ordinary KLM space is already zero by the rank formula
$\sum_k\rk(g_k-I)-\dim\ker(\lambda\prod_k g_k-I)$. This accounts
for every irreducible summand of $D_*$.

It follows that $J=\KLM_\lambda(D_*)$ and $J^\perp=M_*$. The canonical
projection on $M_*$ is induced by $\pi_*^{\oplus n}$. This proves
the lattice formula without requiring a splitting over $\mathcal O$.
The signature follows from the KLM eigenangle formula
\cite[Thm.~1.1]{Neg} applied to $V_*$.
\end{proof}

For an irreducible $F_n$-representation with nonzero KLM image, the entire KLM representation descends if all the conditions in
\eqref{eq:good} hold. If any condition fails, the output is zero.
For reducible inputs, intermediate ranks arise when only some
irreducible summands survive. The construction
extends the marked input and output groups; its pullback to the free
group is always a subquotient of ordinary KLM.

For example take $d_i=d$ for all $i$, with $d$ an odd prime
larger than $n+1$ and $n\ge2$. For the character
$g_i=\lambda=e^{2\pi i/d}$ and trivial input braid action, $\mathcal I_{\ne 1}$ is empty.
The sum of the $n$ local eigenangles and the parameter angle is
$(n+1)/d<1$. The rank-one signature formula gives signature $(n,0)$.
The full braid group acts, and each middle meridian has order exactly
$d$: its nontrivial eigenvalue is $\lambda^2$ and the other eigenvalues
are one. This gives an infinite family with nontrivial finite-order
meridian images on a positive lattice. Removing all order relations
gives the ordinary KLM action on the same marking for any input representation satisfying the hypotheses.

Resonance matters even when the parameter has the prescribed order.
Take $(d_1,d_2)=(4,\infty)$, $\lambda=i$, $g_1=-i$ and $g_2=i$ on
a one-dimensional representation. Its KLM representation has dimension
two and signature $(1,1)$ before imposing the order relation. The first output meridian is
$T_1=\left(\begin{smallmatrix}1&i-1\\0&1\end{smallmatrix}\right)$.
Thus $T_1^4-I$ has image the first coordinate line. By
\eqref{eq:propagate}, $(T_2-I)$ sends that line onto the second, so
the output for $C_4*\Z$ is zero. Here $\lambda^4=1$; it is the
resonant eigenspace in \eqref{eq:good} that detects this loss.

\begin{example}[A proper reduction with its integral form]\label{ex:proper}
Let $(d_1,d_2,d_3)=(2,8,8)$, $s=e^{2\pi i/16}$, $\xi=s^2$,
$K=\Q(s)$, $\mathcal O=\Z[s]$ and $\lambda=\xi$. Let
$B=\langle\sigma_2\rangle$ interchange the two order-eight meridians.
On $K^2$ with its standard form take
\[
       g_1=\diag(1,-1),\qquad g_2=g_3=\xi I,\qquad
       \sigma_2\longmapsto I,\qquad
       A=\{(u,v)\in\mathcal O^2:u-v\in 2\mathcal O\}.
\]
The input operators respect the Artin action and preserve $A$ along with their inverses. The KLM representations of the two input characters have dimensions two and three: the first has one local kernel and neither has an infinity
kernel. The angle sums including the parameter are $3/8$ and $7/8$,
so the KLM form before imposing the order relations has signature $(5,0)$. Since $\mathcal I_{\ne 1}=\{1\}$, the
theorem selects the first character and the quotient has signature
$(2,0)$. In its two surviving coordinate directions the complete
action of $(C_2*C_8*C_8)\rtimes\langle\sigma_2\rangle$ is
\begin{equation}\label{eq:exaction}
 T_1=I,\quad T_2=\begin{pmatrix}\xi^2&\xi-1\\0&1\end{pmatrix},
 \quad T_3=\begin{pmatrix}1&0\\\xi(\xi-1)&\xi^2\end{pmatrix},
 \quad S_2=\begin{pmatrix}0&\xi\\1&1-\xi\end{pmatrix}.
\end{equation}
The order-eight generators act with order four and do not commute.
The matrix $S_2$ conjugates $T_2$ to $T_3$, and conjugates $T_3$
to $T_3^{-1}T_2T_3$, giving the Artin action on the output.
The KLM Hermitian matrix is $|\xi-1|^2 G$, where
\[
 G=\begin{pmatrix}s+s^{-1}&s^{-1}\\s&s+s^{-1}\end{pmatrix}.
\]
Its eigenvalues after removing that positive scalar are
$2\cos(\pi/8)\pm1$. All four matrices preserve this integral form.
Projection of $A$ onto the first coordinate line gives $\mathcal O$, whereas
$A\cap V_*=2\mathcal O$ in the same coordinate. The actual output
lattice is $\mathcal O^2$. Using the intersection lattice as input gives the
marked sublattice $(2\mathcal O)^2$, of index $2^{16}$ since
$[K:\Q]=8$. This comparison uses the same ambient coordinates
and the same Hermitian form on both lattices. Multiplication by $2$
is an equivariant similarity of multiplier $4$ between them.
\end{example}

\subsection{Comparison with the peripheral homology lattice}
There are two operations here: imposing relators on the coefficient
homology of the punctured fiber, and imposing them on the output
meridian action. The following observation identifies the first one
for cyclic orders. Let $\Delta_* = F_n*\langle z\rangle$ and
$\Delta_{\mathbf d}=\Gamma_{\mathbf d}*\langle z\rangle$, acting
on $A$ as in Theorem~\ref{th:integral}. On both sides mark the
finite meridians $x_i$, $z$ and the inverse boundary product, using
the same order of factors, by maps from infinite cyclic groups.
These give the usual peripheral images on the free side.

\begin{proposition}\label{prop:fill}
The quotient $\Delta_*\to\Delta_{\mathbf d}$ induces an isomorphism
of the saturated peripheral middle lattices $M_{\mathcal P}$ of
Definition~\ref{def:peripheral}.
\end{proposition}
\begin{proof}
Degree-one cycles are unchanged by adding the two-cells $x_i^{d_i}$.
The added boundary at $i$ has only $i$th coordinate nonzero, equal to
$(I+g_i+\cdots+g_i^{d_i-1})a$, and this lies in $\ker(g_i-I)$.
It is already the cycle of the $i$th marked peripheral circle. Thus
the map on homology is onto and its kernel lies in the original
boundary image. All marked circle chains and their coefficient
actions map to the corresponding ones on the quotient, so the
peripheral image maps onto the new peripheral image. Taking quotients
and then removing torsion proves the assertion over $\mathcal O$.
For a finite cyclic factor its subgroup homology has the same image,
since the circle homology surjects onto it.
\end{proof}

In particular this fiber calculation alone would retain all five
dimensions in Example~\ref{ex:proper}. Imposing the relations on the
output action is what gives the rank-two representation. For an
arbitrary group no identification of these two operations is assumed.

\section{Commuting meridians and Artin relations}\label{sec:graph}
\begin{definition}[Graph product and subspaces of the input representation]\label{def:graph}
Let $\mathcal E$ be a simple graph on the $n$ marked vertices. Give
vertex $i$ the order $d_i\in\{2,3,\ldots,\infty\}$. The graph product
of the resulting cyclic groups has presentation
\[
 \Gamma(\mathcal E,\mathbf d)=
 \langle x_1,\ldots,x_n \mid x_i^{d_i}=1\ (d_i<\infty),\
              x_i x_j=x_j x_i\ (\{i,j\}\in\mathcal E)\rangle .
\]
Choose $B$ as in Definition~\ref{def:marked} so that its Artin
action preserves the normal closure of these relations and its
peripheral permutation preserves the vertex orders. Use the field,
unitary input representation $V$, stable lattice $A$ and parameter of
Section~\ref{sec:quot}. First form the subspace $V_*$ of
\eqref{eq:good} using just the vertex orders.
For a subspace $U\subset V_*$ let $\operatorname{core}(U)$ denote its
largest $F_n$-stable subspace. It is computed by repeatedly intersecting
$U$ with its translates by $g_k^{\pm1}$ until the dimension stabilizes.
All kernels and images in the next formulas are taken in $V_*$.
Set $D_{ij}=(g_i-I)(g_j-I)$ for each edge and define
\begin{align}
 V_{\rm sep}&=\operatorname{core}\Bigl(\bigcap_{\{i,j\}\in\mathcal E}
                              \ker D_{ij}\Bigr),\label{eq:graphsep}\\
 Z_{ij}&=\operatorname{core}\Bigl(
     \im D_{ij}\ \cap\ \ker(\lambda g_i g_j-I)\
     \cap\!\bigcap_{k\ne i,j}\ker(g_k-I)\Bigr),\label{eq:graphex}\\
 V_{\mathcal E}&=V_{\rm sep}\ \oplus\
                       \bigoplus_{\{i,j\}\in\mathcal E}Z_{ij}.
 \label{eq:graphkeep}
\end{align}
\end{definition}

If all vertex orders are infinite, $\Gamma(\mathcal E,\mathbf d)$ is a
right-angled Artin group. On each edge the two input operators commute
and are unitary. The image of $D_{ij}$ is the sum of the joint
eigenspaces on which both eigenvalues differ from one. Thus $Z_{ij}$
is orthogonal to $\ker D_{ij}$ and is fixed pointwise by every $g_k$
with $k\notin\{i,j\}$. It follows that the sum in \eqref{eq:graphkeep}
is orthogonal.

\begin{theorem}[Graph products of cyclic groups]\label{th:graph}
Let $\pi_{\mathcal E}$ denote the input orthogonal projection onto
$V_{\mathcal E}$. For the full set of graph and order relations, the
output of Theorem~\ref{th:descent} is canonically
\[
  Q=\KLM_\lambda(V_{\mathcal E}),\qquad
  L_Q=\im\bigl((\pi_{\mathcal E}A)^n\longrightarrow Q\bigr),\qquad
  h_Q=h_{\KLM,V_{\mathcal E}}
\]
up to isometry. The acting group is $\Gamma(\mathcal E,\mathbf d)
\rtimes B$. Its marked generators and all its prescribed braid
symmetries act by the KLM matrices on $\KLM_\lambda(V_{\mathcal E})$.
The rank and positive-definiteness test are those of
Theorem~\ref{th:cyclic} with $V_{\mathcal E}$ in place of $V_*$.
The construction recovers the full original KLM action when the graph
has no edges and all $d_i=\infty$.
\end{theorem}

\begin{proof}
Fix an edge and an irreducible complex constituent $Y$ of $V_*$.
We determine when the two output meridians associated with this edge
commute on $\KLM_\lambda(Y)$. The constituents form an orthogonal
direct sum because the input representation is unitary at the chosen
embedding. The subspaces in \eqref{eq:graphsep}--\eqref{eq:graphkeep} are
defined over $K$ by finite systems of linear equations, so their
formation commutes with scalar extension. We show that they select
precisely the constituents satisfying all the edge relations.

Write $T_i$ for the free output meridians and $\iota_i$ for the
coordinate maps to $\KLM_\lambda(Y)$. For $i<j$, multiplication of
the ambient rows gives
\[
 (T_iT_j-T_jT_i)w=
       E_i(g_j-I)Q_jw-\lambda E_j(g_i-I)Q_iw,
       \qquad T_k-I=E_k Q_k .
\]
If $D_{ij}=0$, the two displayed terms lie in their local coordinate
kernels, so the output operators commute.

Assume $D_{ij}\ne0$ and take a joint eigenvector $v$ with eigenvalues
$a,b\ne1$ at $i,j$. The ordinary middle space is nonzero: its rank
is $\sum_k\rk(g_k-I)-\dim\ker(\lambda P-I)$, $P=\prod_k g_k$.
Here $\ker(\lambda P-I)\subseteq\im(P-I)\subseteq\sum_k\im(g_k-I)$,
by $\lambda\ne1$ and telescoping the product. The two images at $i,j$
share $v$, so the last sum has dimension strictly less than the sum
of the ranks. Consequently
\[
   \{u\in Y:\iota_k u=0\text{ for every }k\}=0.
\]
Indeed, this subspace is $F_n$-stable by \eqref{eq:propagate}. It is
proper because the images of the coordinate maps span the nonzero
middle space. Irreducibility of $Y$ therefore forces it to be zero. The same equations show
that $\iota_k u=0$ implies $(g_k-I)u=0$.

On the span of $E_i v,E_j v$, the two meridians have matrices
\begin{equation}\label{eq:edgerows}
 U=\begin{pmatrix}\lambda a&b-1\\0&1\end{pmatrix},\qquad
 W=\begin{pmatrix}1&0\\\lambda(a-1)&\lambda b\end{pmatrix}.
\end{equation}
The determinant of $UW-WU$ is
$\lambda(a-1)(b-1)(\lambda-1)(\lambda ab-1)$. If the output
operators commute and $\lambda ab\ne1$, both coordinate vectors
map to zero, contradicting $(g_i-I)v\ne0$. Thus $\lambda ab=1$.
In this case the image of $UW-WU$ is exactly the line spanned by
$(b,1)^t$. Hence $b\iota_i v+\iota_jv=0$.
For each $k\ne i,j$ apply $T_k-I$ to this equality. By
\eqref{eq:propagate} it gives a scalar multiple of $\iota_kv$.
The scalar is $ab-1$, $b(1-\lambda)$, or $\lambda(ab-1)$,
according as $k<i$, $i<k<j$, or $k>j$. Each is nonzero.
It follows that $\iota_kv=0$ and hence $g_kv=v$ for every such $k$.
The line through $v$ is now $F_n$-stable, so irreducibility implies
$Y=\C v$. Conversely a character supported on exactly these two
vertices with $\lambda ab=1$ has middle dimension $2-1=1$. Its
output satisfies every commuting relation. This proves the exceptional
case in \eqref{eq:graphex}, including for any positions of the marked generators in the ordered tuple.

For a constituent failing an edge test the relation acts nontrivially
on its nonzero, $F_n$-irreducible KLM space, by the irreducibility
argument in Section~\ref{sec:quot}.
The span of the relator differences therefore contains the whole
middle summand. All other constituents survive. Inside $V_*$,
$V_{\mathcal E}$ is equivalently the largest submodule whose KLM action factors through the quotient by the full normal
subgroup of relators. This also proves its $B$-stability:
the KLM operator of $b\in B$ transports the middle space of any
$F_n$-submodule $Y'$ onto that of $bY'$, by the Artin action on the
augmentation ideal and on coefficients. The normal subgroup is
$B$-stable. The orthogonal splitting gives the claimed form; the
blockwise projection of the input lattice gives the claimed lattice,
as in \eqref{eq:cyclic}.
\end{proof}

\begin{example}[A positive rank-three action of $(\Z^2*\Z)\rtimes
 \langle\sigma_1\rangle$]\label{ex:graph}
Take the sole edge $\{1,2\}$, infinite vertex orders, $s=e^{2\pi i/16}$,
$\xi=s^2=\lambda$ and $\mathcal O=\Z[s]$. On a two-dimensional representation
with standard lattice choose
\[
 g_1=\diag(\xi,1),\qquad g_2=\diag(1,\xi),\qquad
 g_3=\begin{pmatrix}0&1\\1&0\end{pmatrix},\qquad
 \sigma_1\longmapsto g_3 .
\]
The swap exchanges $g_1,g_2$ and commutes with $g_3$, as required
by the Artin action. This representation is irreducible and $D_{12}=0$. Adjoin
orthogonally a one-dimensional representation with all three $g_i=\xi$, trivial braid
action and lattice $\mathcal O$. The extra line fails the edge test:
all three generators act nontrivially, and its KLM representation
has dimension three. On the two-dimensional summand the three
local ranks are one and $\lambda g_1g_2g_3=\xi^2g_3$ has no fixed
vector. Thus the KLM representation of the full input has dimension six,
and the graph quotient has dimension three.

On the three copies of the two-dimensional summand, take the first coordinate, the second coordinate, and the difference of the two coordinates, respectively. They
identify the projected middle lattice with $\mathcal O^3$. Writing the
resulting coordinates as $(u,v,w)$, the complete action is
\begin{align*}
 T_1(u,v,w)&=(\xi^2u-w,v,w),&
 T_2(u,v,w)&=(u,\xi^2v+w,w),\\
 T_3(u,v,w)&=(u,v,\xi(\xi-1)(u-v)-\xi w),&
 S_1(u,v,w)&=(v,u,-w).
\end{align*}
Thus $T_1,T_2$ commute, $T_1,T_3$ do not, and $S_1$ interchanges
$T_1,T_2$ while centralizing $T_3$. Put $c=s+s^{-1}$,
$d=|\xi-1|^2$ and $z=s^{-1}(1-\bar\xi)$. The KLM block formula
gives the following Hermitian matrix on this lattice:
\begin{equation}\label{eq:graph-hermitian-matrix}
 G=\begin{pmatrix}dc&0&z\\0&dc&-z\\\bar z&-\bar z&c\end{pmatrix}.
\end{equation}
It is integral. Since $|z|^2=d$, the last Schur complement is
$c-2/c>0$ with $c=2\cos(\pi/8)$. This verifies positive definiteness
directly on the displayed basis.
The ancillary source file \path{anc/verify_marked_relations.py}
checks the quotient maps, the identity for the integral Hermitian matrix, and the edge
computations by exact arithmetic. From the source directory run
\texttt{python anc/verify\_marked\_relations.py} with Python 3 and SymPy.
\end{example}

\subsection{Alternating words for character inputs}
For an Artin edge of length $m\ge2$ write $\mathcal A_m(U,W)$ for the
alternating word of length $m$ starting with $U$. Rank-one input
meridians on this edge have values $a,b$. An odd edge already forces
$a=b$ on the input. The following identities evaluate the output
relation and also hold when $\lambda g_1\cdots g_n$ has eigenvalue one. We use $U,W$ below
for the two full ambient meridian matrices.

\begin{proposition}\label{prop:artin}
The difference $\mathcal A_m(U,W)-\mathcal A_m(W,U)$ is
\begin{align*}
 &\lambda^{k-1}\Bigl(\sum_{r=0}^{k-1}(ab)^r\Bigr)(UW-WU)
       &&(m=2k),\\
 &\lambda^k\Bigl(\sum_{r=0}^{2k}(-a)^r\Bigr)(U-W)
       &&(m=2k+1,\ a=b).
\end{align*}
The same identities hold on the middle space. For an even edge the
relation holds if the displayed scalar vanishes or the two middle
meridians commute. For an odd edge it holds if the scalar vanishes
or the two middle meridians agree. These tests are necessary and
sufficient. In the odd case the equality alternative holds precisely
when $a=b=1$ or when the character is supported on exactly the two
vertices with $\lambda a^2=1$ (with $a=b\ne1$).
\end{proposition}
\begin{proof}
On the two specified coordinate lines the matrices are \eqref{eq:edgerows}.
The products $UW,WU$ have common trace $\lambda(1+ab)$ and
determinant $\lambda^2ab$. Cayley--Hamilton gives the first identity
from the initial differences $0,UW-WU$ at $k=0,1$. For the odd
words the starting differences are $U-W$ and
$\lambda(a^2-a+1)(U-W)$ by direct multiplication. The same recurrence
with $b=a$ gives the stated alternating sum.
The same identities hold on the full space: generically the
complement $\ker Q_i\cap\ker Q_j$ is fixed by both meridians, since
the restriction of the two rows $Q_i,Q_j$ to the two coordinate lines
has determinant $(\lambda-1)(\lambda ab-1)$. Polynomial continuation
in the character parameters and $\lambda$ then gives the identities
also on the exceptional locus (on $a=b$ for odd words).
For an odd edge, equality of the two middle operators implies
commutation. The single-edge proof above therefore gives the alternatives
listed for $a=b$. Both trivial inputs have trivial output meridians;
on the exceptional middle line both act by $\lambda a$.
\end{proof}

For example, if $m=3$ and $a=b=e^{\pi i/3}$ then $a^2-a+1=0$.
Taking $n=2$ and $\lambda=a$ gives a positive rank-two representation
of the three-strand braid group, with two noncommuting meridians
satisfying the length-three relation. For the braid relation between two complex reflections in dimension
two, see Shen \cite[Prop.~3.14 and its introduction]{Shen}, who
traces the criterion to work of Coxeter and Kilmoyer. Here the specialization is expressed in the
input characters for KLM and includes the extra possibilities produced by
the radical quotient.

\section{Different relations on the input and the output}\label{sec:transfer}
The input and output groups may be quotients of $F_n$ by independently
chosen normal subgroups. Work with the ordered meridians and coefficient
ring of Section~\ref{sec:quot}. Put $R_F=\mathcal O[F_n]$. Form the matrices
$\mathcal H,\mathcal T(x_i)$ of Definition~\ref{def:inputideal} over
$R_F$ and extend $\mathcal T$ linearly to the algebra homomorphism
$R_F\to M_n(R_F)$. A group relator $w=1$ corresponds to $f=w-1$;
a polynomial in the marked operators is any element $f\in R_F$.
\begin{definition}[Transfer of ideals]\label{def:transfer}
For an input ideal $\mathfrak a$ and an output ideal $\mathfrak b$, both two-sided in $R_F$, define
\begin{align*}
 P_\lambda(\mathfrak b)&=\big\langle [\mathcal H\mathcal T(f)]_{ij}:
                  f\in \mathfrak b,\ 1\le i,j\le n\big\rangle_{\rm two\text{-}sided},\\
 D_\lambda(\mathfrak a)&=\{f\in R_F:\mathcal H\mathcal T(f)\in M_n(\mathfrak a)\}.
\end{align*}
\end{definition}

\begin{theorem}[Transfer of input and output identities]\label{th:transfer}
The maps $P_\lambda$ and $D_\lambda$ form an order-preserving
adjunction on the poset of two-sided ideals. This adjunction restricts
to the subposet of involution-stable ideals. In either poset,
\[
       P_\lambda(\mathfrak b)\subseteq \mathfrak a\quad\Longleftrightarrow\quad
       \mathfrak b\subseteq D_\lambda(\mathfrak a).
\]
In particular $P_\lambda$ preserves sums and $D_\lambda$ preserves
intersections. Generators of $\mathfrak b$ as a two-sided ideal suffice in the
formula for $P_\lambda$. For any input representation $V$ with a positive definite invariant
Hermitian form and a stable full lattice $A$, let $\mathfrak a_V$ be the kernel of $R_F\to\operatorname{End}_K(V)$.
Then the annihilator $\mathfrak a_{\KLM_\lambda(V)}$ in $R_F$ of the middle
representation equals $D_\lambda(\mathfrak a_V)$. For every prescribed output ideal $\mathfrak b$, this gives
\[
        \mathfrak b\subseteq \mathfrak a_{\KLM_\lambda(V)}
        \quad\Longleftrightarrow\quad P_\lambda(\mathfrak b)\subseteq \mathfrak a_V .
\]
For a prescribed output ideal $\mathfrak b$, the largest submodule of $V$ whose
middle representation is killed by $\mathfrak b$ is
$Y=\operatorname{Ann}_V P_\lambda(\mathfrak b)$, the common kernel in $V$ of the
elements of $P_\lambda(\mathfrak b)$. On $\KLM_\lambda(Y)$ use the projected
lattice $L_{\rm proj}=q_Y((\pi_Y A)^n)$ of Definition~\ref{def:projection}.
Then the $\mathfrak b$-annihilated subspace of $\KLM_\lambda(V)$ is
$\KLM_\lambda(Y)$ with its ordinary middle form, and orthogonal
projection of the full middle lattice onto this subspace gives
$L_{\rm proj}$. If a group $B$ as in Definition~\ref{def:marked} acts
compatibly on $V$ and its Artin action preserves $\mathfrak b$, this description
is equivariant for that group.
\end{theorem}
\begin{proof}
Write $E(f)=\mathcal H\mathcal T(f)$. For words $u,v\in F_n$,
invariance of $\mathcal H$ gives
\[
 E(ufv)=(\mathcal T(u)^{-1})^*E(f)\mathcal T(v),
 \qquad E(f^*)=E(f)^* .
\]
Linearity gives the ideal and generator assertions, and the second
identity gives compatibility with involution. The stated equivalence
follows by testing every entry; testing sums or intersections gives
the two preservation laws.

On $V^n$ the kernel of the evaluated $\mathcal H$ is precisely the
middle kernel. An element $f$ annihilates the middle space if and
only if the evaluated $\mathcal T(f)$ has image in this kernel, or
equivalently every entry of $E(f)$ belongs to $\mathfrak a_V$. This proves the
annihilator identity. Extend scalars to $\C$ as in the proof of
Theorem~\ref{th:ideal} and decompose the input orthogonally into
irreducibles. Each middle image is zero or irreducible by the two
cases in Section~\ref{sec:quot}, and its $\mathfrak b$-annihilated subspace
is therefore zero or the whole middle summand. Summing the surviving
summands and projecting the marked lattice gives the assertion just
as in Theorem~\ref{th:ideal}. Braid equivariance follows by transport
of submodules and stability of $\mathfrak b$.
\end{proof}

In particular, take two arbitrary marked quotients
$\Gamma_{\rm in}=F_n/N_{\rm in}$ and $\Gamma_{\rm out}=F_n/N_{\rm out}$.
Use an input lattice over $\mathcal O[\Gamma_{\rm in}]$ and take
$\mathfrak b=\langle w-1:w\in N_{\rm out}\rangle$ in $R_F$. Only the image of
$P_\lambda(\mathfrak b)$ in $\mathcal O[\Gamma_{\rm in}]$ acts on $V$.
The output lattice carries a representation of $\Gamma_{\rm out}$.
If the input also carries a compatible action of $B$ as in
Definition~\ref{def:marked} and its Artin action preserves both
normal subgroups, the KLM operators extend the
output action to $\Gamma_{\rm out}\rtimes B$.
No inclusion between $N_{\rm in}$ and $N_{\rm out}$ is required. When the two normal subgroups coincide, we recover
Theorem~\ref{th:ideal}. The formula for the annihilator can also be iterated along a sequence of unitary input representations with positive definite metrics by composing the maps $D_\lambda$ in the order of convolution.

\paragraph{Other symmetry groups.}
Write $\rho_Q$ for the resulting action of $\Gamma_{\rm out}$. Let
$G$ be any group and $\alpha:G\to\operatorname{Aut}(\Gamma_{\rm out})$
an action. By the defining relations of the semidirect product,
$\rho_Q$ extends to an action of $\Gamma_{\rm out}\rtimes_\alpha G$
preserving $L_Q$ and $h_Q$ if and only if there is a homomorphism
$g\mapsto P_g$ to the group of $\mathcal O$-linear isometries of $L_Q$
such that
\[
  P_g\,\rho_Q(x_i)\,P_g^{-1}=\rho_Q(\alpha_g(x_i))
  \qquad (1\le i\le n).
\]
For instance, if $\alpha_g$ is conjugation by $\theta(g)$ for a
homomorphism $\theta:G\to\Gamma_{\rm out}$, take $P_g=\rho_Q(\theta(g))$.
For an arbitrary action on the marking, the existence of these
isometries is an additional condition. Take $\Gamma_{\rm in}=
\Gamma_{\rm out}=F_1$, the input character $x\mapsto-1$, and
$\lambda=i$ over $\Q(\zeta_8)$. Inversion $x\mapsto x^{-1}$ fixes
the input character. The middle line has $T=-i$ (the middle kernel
is zero), whereas $T^{-1}=i$. No linear isometry of the line
conjugates $T$ to $T^{-1}$.

\subsection{Varying the parameter}
Fix a complex irreducible unitary input representation $Y$ and a finite list of output
relations $f_1,\ldots,f_k$ with coefficients independent of the
parameter.

\begin{definition}[Laurent polynomial of the prescribed relations]\label{def:parameter}
Set $s^2=t$ and replace $\lambda$ by $t$ in $\mathcal T(x_i)$ and
$\mathcal H$. The entries of $s\mathcal H$
now involve only $t$ and the marked group elements. Let $p_Y(t)$ be
the gcd of all evaluated entries of $(s\mathcal H)\mathcal T(f_j)$
in $\C[t,t^{-1}]$, up to Laurent units, with gcd zero for a list of
zeros (including an empty list).
\end{definition}
At any $|\lambda|=1$, $\lambda\ne1$,
all target identities hold on $\KLM_\lambda(Y)$ exactly when
$p_Y(\lambda)=0$. Indeed a common zero in this one-variable Laurent
polynomial ring is precisely a zero of the gcd. To add a relation, take the gcd of the current polynomial and the
polynomials for the new relation. Theorem~\ref{th:ideal} then gives
the middle dimension and signature for each selected subrepresentation.

For example put $a=e^{2\pi i/5}$ and use the three input characters
$(a,1),(1,a),(a,a)$ with target relation $x_1x_2=x_2x_1$. On the
first two the relation holds identically. For the last, multiplication
of the two rows gives, up to a Laurent unit,
$p_Y(t)=(t-1)(a^2t-1)$. Using $x_1x_2-x_2x_1$ in this computation
gives the same ideal as the group commutator minus one.
With $\lambda=e^{2\pi i \ell}$ and $s=e^{\pi i \ell}$, the signature is
$(2,0)$ for $0<\ell<4/5$ except at $\ell=3/5$, where it is $(3,0)$.
At $\ell=4/5$ the rank is zero; for $4/5<\ell<1$ the signature is $(0,2)$.
At the first exceptional point an additional character satisfies the
target relation. At the second point the first two middle spaces
vanish.

\subsection{Finite real reflection groups}
\begin{example}[A cyclic input and a dihedral output]\label{ex:dihedral}
For $m\ge3$ take $s=e^{\pi i/m}$, $\lambda=s^2$, $a=-s^{-2}$ and
$\mathcal O=\Z[s]$. On a single copy of $\mathcal O$ let both marked
generators act by $a$, so that their image is cyclic. In the two
convolution coordinates use the basis change $\diag(-s^{-1},1)$.
With $b=s+s^{-1}$ the transformed rows and Hermitian matrix are
\[
 U=\begin{pmatrix}-1&b\\0&1\end{pmatrix},\qquad
 W=\begin{pmatrix}1&0\\b&-1\end{pmatrix},\qquad
 G=b\begin{pmatrix}2&-b\\-b&2\end{pmatrix}.
\]
These matrices follow by evaluating $\mathcal T(x_1),\mathcal T(x_2)$
and $\mathcal H$ at the given character and parameter. Since $0<b<2$, the middle space has the full lattice
$\mathcal O^2$ and $G$ is positive. Both matrices square to one.
The eigenvalues of $UW$ are $s^2,s^{-2}$, so it has exact order $m$.
The two involutions generate the dihedral group of order $2m$:
the $m$ rotations and their products with $U$ are disjoint by
determinant. These are the standard geometric simple reflections of
$I_2(m)$ \cite[\S5.3]{Hum}. The source identity $x_1=x_2$ fails on
the output and the target identity $x_1^2=1$ fails on the input.
\end{example}

\begin{corollary}[Rank reduction at $\mu=e^{2\pi i/h}$]\label{cor:coxwall}
Let $W$ be a finite irreducible real Coxeter group of rank $d\ge2$
with Coxeter number $h$. Regard its reflection representation $E$
as an $F_d$-module through its ordered simple reflections. With
$\mu=e^{2\pi i/h}$ and its square root $e^{\pi i/h}$,
$Y=\KLM_\mu(E)$ has rank $d-1$, a positive definite middle form, and
a stable full lattice. Convolving $Y$ with parameter $\mu^{-1}$
recovers the marked representation $E$ over the coefficient field.
For a suitable choice of square root the recovered invariant form
is positive definite.
\end{corollary}
\begin{proof}
Work over a conjugation-stable cyclotomic field containing the
simple-root matrix entries $2\cos(\pi/m_{ij})$ and $\mu^{1/2}$.
The lattice spanned by simple roots over its integers is stable,
and the standard metric is positive. The ordered product $C$ of
the simple reflections has eigenangles $m_j/h$ for the exponents
$m_j$. They lie in $(0,1)$, have sum $d/2$, and the largest is
$(h-1)/h$ with multiplicity one \cite[\S\S3.16--3.20]{Hum}.
Multiplying $C$ by $\mu$ moves precisely this last angle to zero.
Thus $\theta(\mu C)=d/2+d/h-1$ and $\dim\ker(\mu C-I)=1$.
Each simple reflection has moving rank one and eigenangle sum
$1/2$. Theorem~\ref{th:ideal} with no relations gives $R_0=d$,
$\kappa=r=1$ and signature $(d-1,0)$. The lattice is the image of the
input lattice in the middle space. Since $E$ is irreducible and
has dimension at least two, there is no nonzero $F_d$-map to or from
an almost trivial character. Thus $E\in\mathrm{Mod}^{\rm NT}$, and
the KLM inverse theorem \cite[Thm.~5.7]{HN} gives
$\KLM_{\mu^{-1}}(Y)\simeq E$ over the field. By absolute irreducibility
its nondegenerate invariant Hermitian form is a real scalar multiple
of the positive reflection metric; switching the square root selects
the positive sign.
\end{proof}

Rank reduction from finite reflection tuples by middle convolution
also appears in \cite{DRJ,Vay}. In Corollary~\ref{cor:coxwall} the Coxeter number $h$ fixes the
parameter $\mu=e^{2\pi i/h}$. This gives a positive form and a stable
lattice in each type.

\begin{proposition}[The Hermitian matrix in reflection coordinates]
\label{prop:reflectioncoords}
Let $(E,h_0)$ have positive metric, linear in the second argument.
Take any ordered tuple of reflections
$g_i v=v-\alpha_i h_0(\alpha_i,v)$ with
$h_0(\alpha_i,\alpha_i)=2$, and set $b_{ij}=h_0(\alpha_i,\alpha_j)$.
Write $c=s+s^{-1}$ and use the $n$ coordinates
$z_i=h_0(\alpha_i,v_i)$ on the quotient of $E^n$ by the local
kernels. In these coordinates the middle form is represented by the Hermitian matrix
\[
 G_{ii}=c,\qquad G_{ij}=s^{-1}b_{ij}\quad(i<j),
 \qquad G_{ji}=\overline{G_{ij}} .
\]
The middle space is $\C^n/\ker G$. The $i$th meridian leaves the
other coordinates fixed and sends
\[
 z_i\longmapsto-\lambda z_i-\lambda\sum_{j<i}b_{ij}z_j
                              -\sum_{j>i}b_{ij}z_j .
\]
For data over the coefficient field and a stable full lattice $A$,
the marked lattice is the image of
$\bigoplus_i h_0(\alpha_i,A)$ in the quotient over that field.

Suppose the graph of nonzero off-diagonal pairings is a tree with
$n\ge2$. Let $A_{\rm wt}$ be its weighted adjacency matrix, with entries $|b_{ij}|$ on the edges and zero elsewhere,
including the diagonal. The middle quotient is nonzero and positive
definite exactly when $c\ge\rho(A_{\rm wt})$, the spectral radius.
Its rank is $n$ for $c>\rho(A_{\rm wt})$ and $n-1$ for $c=\rho(A_{\rm wt})$.
No finiteness condition on the group generated by the reflections
is needed.
\end{proposition}
\begin{proof}
Substitute $g_i-I=-\alpha_i h_0(\alpha_i,-)$ in the formulas for $\mathcal T(x_i)$
and $\mathcal H$. The map $(v_i)\mapsto(z_i)$ is onto and its kernel is
exactly the sum of the local kernels. The evaluated form pulls back
from the displayed $G$, and the image of $A^n$ in these coordinates
is the direct sum stated above. On a tree choose diagonal phases
recursively from a root so that every off-diagonal entry of the
transformed matrix is $-|b_{ij}|$. The resulting matrix is $cI-A_{\rm wt}$.
It is positive semidefinite precisely when $c\ge\rho(A_{\rm wt})$, and
Perron--Frobenius gives a simple kernel at equality. A negative
eigenvalue survives quotient by the radical. For instance the path
$A_3$ has edge weights one and $\rho(A_{\rm wt})=\sqrt2$. At $\lambda=i$,
$s=e^{\pi i/4}$, the Hermitian matrix in these coordinates has eigenvalues $0,\sqrt2,2\sqrt2$, giving a positive rank-two input for the
construction of the standard reflection representation of $S_4$.
\end{proof}

\subsection{The MC gate construction in a common computational basis}\label{sec:gates}
\begin{definition}[Middle convolution gate construction]\label{def:gate}
Let $(Q,h,\rho)$ be the middle representation obtained from a marked
representation and its lattice by prescribing the output relations.
At a complex embedding where $h$ is positive definite, fix a single
isometry $\iota$ from the resulting Hermitian space to a quantum
register. The gate associated with the marked generator $x_i$ is
$\iota\rho(x_i)\iota^{-1}$, and a word in the marked generators acts through this same
identification.
For $\dim Q=2^k$ the register is $(\C^2)^{\otimes k}$ with its
computational basis.
\end{definition}
Theorem~\ref{th:transfer} tests the identities of the marked operators.
An explicit intertwiner with a specified representation identifies
those operators in the chosen encoding. In the finite-image example below, inverse convolution recovers the
multi-controlled $X$ and $Z$ gates. The forward construction in
Theorem~\ref{th:densegates} starts with one scalar character and
determines the closure of the generated gate group.

\begin{corollary}[Controlled gates from a single type \texorpdfstring{$B$}{B} action]
\label{cor:controlled}
For $k\ge2$, put $d=2^k$ and $\mu=e^{\pi i/d}$. There is a
rank-$(d-1)$ representation $Y$ with a positive definite invariant
Hermitian form and a stable full lattice such that $\KLM_{\mu^{-1}}(Y)$
realizes the following two $k$-qubit gates in one encoding:
\[
       C^{k-1}X,\qquad C^{k-1}Z .
\]
The first $k-1$ qubits are the controls. The operation acts on the last
qubit when every control is in state $|1\rangle$. The two gates are
the images of the last two marked generators.
\end{corollary}
\begin{proof}
Let $e_1,\dots,e_d$ be the orthonormal coordinate vectors of the
type $B_d$ reflection representation. Take the simple reflections
$s_i$ exchanging $e_i,e_{i+1}$ for $i<d$, and $s_d$ changing the
sign of $e_d$. Their product is a signed $d$-cycle with negative
sign product, of order $2d$. Apply Corollary~\ref{cor:coxwall}
with Coxeter number $2d$ and transport its recovered marked
representation to these coordinates. Normalize the positive
invariant form by its common scalar relative to the standard metric.
Identify $e_{j+1}$ with the length-$k$ binary expansion of $j$,
from the most significant qubit to the least significant.
Then $s_{d-1}$ exchanges precisely $|1^{k-1}0\rangle$ and
$|1^{k-1}1\rangle$, while $s_d$ negates only $|1^k\rangle$.
Both statements use this same identification.
\end{proof}

\begin{center}
\begin{tabular}{@{}lccc@{}}
\toprule
Type & Input rank & $s_{d-1}$ & $s_d$\\
\midrule
$B_4$ (two qubits) & 3 & CNOT & CZ\\
$B_8$ (three qubits) & 7 & Toffoli & CCZ\\
\bottomrule
\end{tabular}
\end{center}

The coefficient ring places a further condition on gates that share a
lattice. If a group preserves a full lattice over the integers of a
number field, every word in its generators has algebraic-integer
trace. Indeed the characteristic polynomial has integral coefficients
at each finite prime, as one sees in a basis of the localized lattice.
This test is invariant under simultaneous conjugation.

\begin{proposition}[The dyadic prime for Hadamard and \texorpdfstring{$\mathsf T$}{T}]
\label{prop:dyadic}
Put $\zeta_8=e^{\pi i/4}$, $K=\Q(\zeta_8)$ and
\[
 \mathsf H=\frac{1}{\sqrt2}\begin{pmatrix}1&1\\1&-1\end{pmatrix},
 \qquad \mathsf T=\diag(1,\zeta_8).
\]
Among localizations of $\mathcal O_K=\Z[\zeta_8]$ at sets of
rational primes, a common stable full lattice exists exactly when
the set contains $2$. Over the integers of any number field, the
obstruction persists under a change of basis and even under
independent global phases on the two gates.
\end{proposition}
\begin{proof}
For $\tau=\operatorname{tr}(\mathsf H\mathsf T)=(1-\zeta_8)/\sqrt2$ one has
\[
 N_{K/\Q}(\tau)=\frac{\Phi_8(1)}{N_{K/\Q}(\sqrt2)}
                =\frac{2}{4}=\frac12.
\]
For a set $\Sigma$ of rational primes an element of
$\mathcal O_K[\Sigma^{-1}]$ has norm in $\Z[\Sigma^{-1}]$. Hence a common
lattice forces $2\in\Sigma$. Conversely the entries of $\mathsf H,\mathsf T$ and
their inverses lie in $\Z[\zeta_8,1/2]$, so the standard free module
of rank two over this ring is stable. For arbitrary phases the
absolute square of the product trace remains
$|\tau|^2=1-1/\sqrt2$, of norm $1/2$ in $\Q(\sqrt2)$. If the
rephased product preserved an integral lattice, its trace and the
complex conjugate of its trace would both be algebraic integers,
contradicting this norm.
\end{proof}

The localized ring $\Z[\zeta_8,1/2]=\Z[1/\sqrt2,i]$ also occurs in
the exact Clifford+$\mathsf T$ synthesis theorem of Giles and Selinger
\cite{GS}; Hermitian lattice methods for circuit synthesis are
developed in \cite{KS}. Proposition~\ref{prop:dyadic} explains why
the dyadic denominator is necessary even when one is free to choose
the lattice. Corollary~\ref{cor:controlled} supplies the KLM
realization for the signed permutation gates.

For later use we record the elementary change of basis that puts all
marked operators on one register. For an output chosen by
Theorem~\ref{th:transfer}, the eigenangle tests $R_0-r=2^k$ and $\kappa=r$
of Theorem~\ref{th:ideal} determine from the input when this applies.

\begin{proposition}[A common algebraic encoding]\label{prop:gateencoding}
Let $(Q,h,L)$ be an output of Theorem~\ref{th:transfer} over a
conjugation-stable number field $K$, with $h>0$ at its chosen
embedding and $\dim_K Q=2^k$. There is a finite conjugation-stable
extension $K'/K$ and a single isometry from $Q_{K'}$ to the standard
$k$-qubit space over $K'$. In this encoding all marked operators
are algebraic unitary matrices satisfying the same group and
polynomial identities, and they preserve the transported full
$\mathcal O_{K'}$-lattice.
\end{proposition}
\begin{proof}
Choose a $K$-basis of $Q$ and orthogonalize it for $h$. The squared
lengths are positive real elements of $K$. Adjoin their positive
square roots and divide the orthogonal vectors by their lengths.
Sending this orthonormal basis in a chosen order to the binary
computational basis gives the isometry. The simultaneous change of
coordinates preserves all identities. It transports
$L\otimes_{\mathcal O_K}\mathcal O_{K'}$ to a lattice invariant
under every marked gate and its inverse.
\end{proof}

\subsubsection*{A dense gate group from a cyclic input}
Fix the root $\lambda$ in the upper half-plane determined by
\[
       \lambda+\lambda^{-1}=\frac{\sqrt{13}-1}{2}.
\]
It is a root of $X^4+X^3-X^2+X+1$, hence an algebraic unit.
Writing $\lambda=e^{2\pi i \ell}$ gives $0<\ell<1/4$. This unit has infinite
order: the other conjugate $(-1-\sqrt{13})/2$ of
$\lambda+\lambda^{-1}$ is less than $-2$, whereas all conjugates of
$\zeta+\zeta^{-1}$ for a root of unity $\zeta$ lie in $[-2,2]$.
Use $s=e^{\pi i \ell}$ throughout.

\begin{theorem}[Dense integral meridians from a cyclic character]
\label{th:densegates}
For $n\ge2$, mark $C_{2n}$ by $n$ copies of one generator, acting on
a line by $a=e^{\pi i/n}$. Put $K=\Q(a,s)$ and let $\mathcal O$
be its ring of integers. Multiplicative middle convolution at
$\lambda$ has rank $n$, positive form $H$ at the displayed embedding,
and full stable lattice $\mathcal O^n$. For each pair of distinct marked meridians, the shortest Artin
relation has length $2n$ when $n$ is even and $n$ when $n$ is odd. In particular all pairs
satisfy the length-$2n$ Artin relation. Writing $T_i$ for the
marked meridians and $U(H)=\{g\in\mathrm{GL}_n(\C):g^*Hg=H\}$ for
the unitary group of the form, one has in the ordinary topology
\[
      \overline{\langle T_1,\ldots,T_n\rangle}=U(H).
\]
For $n=2^k$, the ordered convolution coordinates determine one
algebraic encoding in which words in these meridians and their inverses
approximate every $k$-qubit unitary. In the same encoding $T_1$ fixes
every computational basis state except $|1^k\rangle$, on which it acts
by $\lambda a$. For $k\ge2$ this gate entangles the last two qubits
when the first $k-2$ qubits are in state $|1\rangle$ and the last
two are in state $|+\rangle$.
\end{theorem}

\begin{proof}
At the specified embedding $K$ is closed under conjugation because
$\bar a=a^{-1}$ and $\bar s=s^{-1}$. With input metric $1$,
the signature formula of Theorem~\ref{th:ideal} gives $R_0=n$,
$r=0$ and $\kappa=\ell+n/(2n)-(\ell+1/2)=0$: the product at infinity for that
formula is $\lambda a^n=-\lambda$. Thus the form $H$ on $K^n$ is
positive and has zero radical. The rows and their inverses preserve
$\mathcal O^n$ since $a$ and $\lambda$ are units. For each pair
$U=T_i,W=T_j$ with $i<j$, the first diagonal entry of $UW-WU$ on
the two coordinate lines in \eqref{eq:edgerows} is $\lambda(a-1)^2\ne0$.
Thus both $UW-WU$ and $U-W$ are nonzero. By
Proposition~\ref{prop:artin}, an even length $m=2j$ gives equality
exactly when $(a^2)^j=1$, that is, $2n\mid m$. For odd $m$ the
criterion is $(-a)^m=1$, since $-a\ne1$. As $-a=a^{n+1}$, it has
order $2n$ for even $n$ and $n$ for odd $n$. This gives the stated
minimum in both cases.

Put $t=\lambda a$ and use the inner product $\langle x,y\rangle_H
=x^*Hy$. Each $T_i-I$ has image $K e_i$, and $T_i e_i=t e_i$.
Unitarity gives $T_i=I+(t-1)P_i$ for the $H$-orthogonal projection
$P_i$ onto $e_i$. The number $t$ has infinite order. The closed
group $G$ generated by the $T_i$ therefore contains the circle
$\{I+(e^{i\theta}-1)P_i:\theta\in\R\}$ for each $i$.

Let $v_i=e_i/\|e_i\|_H$, $c_{ij}=\langle v_i,v_j\rangle_H$ and
$E_{ij}x=v_i\langle v_j,x\rangle_H$. By the block formula of
Section~\ref{sec:tangent}, $H_{ij}=s^{-1}|a-1|^2$ for $i<j$,
so $c_{ij}\ne0$. The complexified Lie algebra of $G$ contains
$P_i=E_{ii}$ and, for $i\ne j$, contains
\begin{align*}
 [P_i,P_j]&=c_{ij}E_{ij}-\overline{c_{ij}}E_{ji},\\
 [P_i,[P_i,P_j]]+2|c_{ij}|^2P_i
           &=c_{ij}E_{ij}+\overline{c_{ij}}E_{ji}.
\end{align*}
It contains each $E_{ij}$. The $v_i$ form a basis and the covectors
$\langle v_j,\cdot\rangle_H$ form a dual-space basis, so these
operators span $\operatorname{End}(\C^n)$. Hence the real Lie algebra
of $G$ has dimension $n^2$, the dimension of $\mathfrak u(H)$.
The connectedness of $U(H)$ now gives $G=U(H)$.

For $n=2^k$, orthonormalize $e_1,\ldots,e_n$ in that order as in
Proposition~\ref{prop:gateencoding} and assign the resulting vectors,
in order, to the binary strings in decreasing order. The first is
$v_1$, so $P_1=|1^k\rangle\langle1^k|$ in this encoding. When
$k\ge2$, acting on $|1\rangle^{\otimes(k-2)}\otimes|++\rangle$
produces a state of Schmidt rank two on the last pair: the determinant
of its two-qubit coefficient matrix is $(t-1)/4\ne0$.
\end{proof}

The integral lattice is compatible with density at the chosen
embedding. Extend the other embedding of $\Q(\sqrt{13})$ to an
embedding $\sigma:K\hookrightarrow\C$. Since
$\sigma(\lambda)+\sigma(\lambda)^{-1}=(-1-\sqrt{13})/2<-2$,
$\sigma(\lambda)$ is real and has modulus different from one.
The meridian eigenvalue $\sigma(\lambda a)$ also lies off the unit
circle. Thus the conjugate representation admits no positive definite
invariant Hermitian form. A finite image would admit such a form at
every embedding by averaging. Infinitude also follows directly from
\cite[Prop.~2.7]{NegFinite}, since the input character has nontrivial
finite image and $\lambda$ has infinite order. For a root of unity
$\lambda\ne1$, the corresponding finiteness criterion tests definiteness
on irreducible constituents at every embedding \cite[Thm.~2.10]{NegFinite}.
For the role of conjugate forms in the cyclotomic case, see
\cite[Rem.~5.11]{DR00}.

The density is a statement about the marked meridians on each fixed
register. Equivalently they are the point-pushing generators inside
the pure braid group: the character, being equal on all meridians,
extends with trivial action on the braid factor, and
the configuration-space fibration \cite[Thm.~1]{FN} identifies the
group $F_n\rtimes P_n$ with $P_{n+1}$. Its action by KLM isometries
is the one in \cite{HN,NegMC}. This pure braid image contains the dense subgroup
just proved. In \cite{NegGates} the image of the braid group factor on the full
twisted Burau output is projectively nondense. The two statements
concern different inputs and different acting subgroups. The density
argument gives no bound on synthesis length or on the cost of
implementing one marked meridian by local gates.

\begin{example}[An Artin relation selects the dense summand]
\label{ex:densemarked}
Take $n=2^k$, $k\ge1$, and let the same cyclic generator act by
$\diag(a,-1)$ on two orthogonal input lines. Use the full lattice
\[
 A=\{(x,y)\in\mathcal O^2:x-y\in(1+a)\mathcal O\}.
\]
It is stable because $ax+y=a(x-y)+(1+a)y$ and the cyclic generator
has finite order. Impose the length-$2n$ Artin relation on just the
first pair of output meridians. The meridians obtained from the $a$-line
satisfy the relation by Theorem~\ref{th:densegates}. On the other line
every input generator acts by $-1$,
and $\lambda(-1)^n=\lambda\ne1$, so its ambient KLM form has zero
radical. On the two coordinate lines, the $(1,1)$ entry of $UW-WU$
in \eqref{eq:edgerows} is $4\lambda$. The scalar multiplying this
commutator in Proposition~\ref{prop:artin} is $n\lambda^{n-1}\ne0$.
Hence the middle representation obtained from the second line fails
to satisfy the output relation. By
Theorem~\ref{th:transfer} the projected input is precisely the $a$-line.
Its projected lattice is $\mathcal O$ since $(x,x)\in A$ for each $x$,
and its intersection lattice is $(1+a)\mathcal O$. The inclusion is
strict: $N_{\Q(a)/\Q}(1+a)=\Phi_{2n}(-1)=2$, so $1+a$ remains a
nonunit in $\mathcal O$. Thus the marked quotient selects the dense
action with lattice $\mathcal O^n$, whereas using the intersection
would give $(1+a)\mathcal O^n$ in the same coordinates. The two are
equivariantly similar, with metric multiplier $|1+a|^2$. The next
example gives a different similarity class.
\end{example}

\begin{example}[Nonsimilar lattices on an irreducible rank-two output]
\label{ex:nonsimilar}
Let $s=\zeta_{16}=e^{2\pi i/16}$, $\xi=s^2=\lambda$,
$K=\Q(s)$, $\mathcal O=\Z[s]$, and $\delta=1-s$.
Then $\mathcal O/\delta\mathcal O=\mathbb F_2$ since $\Phi_{16}(1)=2$.
On the standard Hermitian space $Y=K^2$ take
\[
        g_1=\diag(\xi,1),\qquad
        g_2=\begin{pmatrix}0&1\\1&0\end{pmatrix}.
\]
On an orthogonal line $Z$ let both meridians act by $-1$. This is
a representation of $C_8*C_2$. Give $Y\perp Z$ the full lattice
\[
 A=\{(x,y,z)\in\mathcal O^3:x+y\equiv z\pmod{\delta}\}.
\]
Modulo $\delta$ the first generator fixes all coordinates, and the
second interchanges $x,y$ and fixes $z$. Thus both preserve $A$.

Prescribe the marked output group $C_4*C_8$, i.e.\ $T_1^4=T_2^8=I$.
On $Y$ the two local ranks are one, and $(\lambda g_1g_2)^2=\xi^3I$
excludes an infinity kernel. Write a vector in the $j$th summand as $(x_j,y_j)$.
The local kernels give quotient coordinates $u=x_1$, $v=x_2-y_2$.
In this basis the two meridians are
\[
 U=\begin{pmatrix}\xi^2&-1\\0&1\end{pmatrix},
 \qquad W=\begin{pmatrix}1&0\\\xi(\xi-1)&-\xi\end{pmatrix}.
\]
Each matrix has two distinct eigenvalues. The orders of $U$ and $W$
are four and eight, respectively. A direct calculation gives
$\det(UW-WU)=\xi(\xi-1)^2(\xi^2+\xi+1)\ne0$. A common invariant
line over $\C$ would lie in its kernel, so the output is absolutely
irreducible. The middle form is the principal submatrix
of \eqref{eq:graph-hermitian-matrix} on indices $1,3$, hence positive definite.
The input representation $Y$ is absolutely irreducible since $g_2$ exchanges
the two eigenlines of $g_1$. On $Z$ the local kernels vanish
and $\lambda(-1)^2\ne1$ excludes an infinity kernel. The first
output meridian there has eigenvalue $-\xi$, of order eight.
Theorem~\ref{th:transfer} therefore selects exactly $Y$.

Projection of $A$ to $Y$ gives $\mathcal O^2$, since we may take
$z=x+y$. Its intersection with $Y$ imposes $x+y\in\delta\mathcal O$.
On the intersection $x$ takes all values in $\mathcal O$, while
$x-y$ takes exactly $\delta\mathcal O$: one has $2\in\delta\mathcal O$
and $(x,y)=(\delta b,0)$ realizes any $\delta b$. As $u$ and $v$ come
from different summands, the two output lattices are
\[
           L_{\rm proj}=\mathcal O^2,\qquad
           L_{\rm int}=\mathcal O\oplus\delta\mathcal O .
\]
In particular their additive index is two.

Even after forgetting the action, these Hermitian lattices are not
similar. A complex-linear similarity $S$ carrying $L_{\rm proj}$ onto
$L_{\rm int}$ would have $S=\diag(1,\delta)B$ for some
$B\in\mathrm{GL}_2(\mathcal O)$. Write $G$ for the common Hermitian matrix
in ambient coordinates. From $S^*GS=bG$ one gets
$b\in K^+=K\cap\R$ by comparing a nonzero diagonal entry, and
$b^2=\det S\,\overline{\det S}$. Since $N_{K/\Q}(\delta)=2$ and
units of this totally imaginary field have norm $1$, taking norms
gives $N_{K/\Q}(b)=2$. But
$N_{K/\Q}(b)=N_{K^+/\Q}(b)^2$ is a rational square. Moreover,
every endomorphism commuting with the marked action is scalar by
absolute irreducibility. No scalar sends $\mathcal O^2$ onto
$\mathcal O\oplus\delta\mathcal O$, so these lattices are also
nonisomorphic as $\mathcal O[C_4*C_8]$-modules.
\end{example}

\begin{galoiscorollary}
In the coordinates $(u,v)$ of Example~\ref{ex:nonsimilar}, normalize the
invariant Hermitian matrix $G$ by $G_{11}=1$. For the embeddings
$\sigma_k:s\mapsto s^k$, indexed up to complex conjugation, the signatures
of $\sigma_k(G)$ are $(2,0)$ for $k=1,7$ and $(1,1)$ for $k=3,5$.
The rational symmetric form
\[
 B(x,y)=\operatorname{Tr}_{K/\Q}(x^*Gy),\qquad x,y\in K^2,
\]
has signature $(12,4)$ on the rational span of either output lattice.
\end{galoiscorollary}
\begin{proof}
Write $G=\bigl(\begin{smallmatrix}1&h\\\bar h&d\end{smallmatrix}\bigr)$.
The $(1,2)$ entries of $U^*GU=G$ and $W^*GW=G$, respectively, give
\[
 h=\frac{1}{1-\xi^2},\qquad
 d=\frac{1}{(1-\xi)(1-\xi^{-1})}.
\]
Using $\xi^4=-1$, we obtain
$\det G=d-h\bar h=(1+\xi+\xi^{-1})/2$.
Under $\sigma_k$ this becomes $(1+2\cos(k\pi/4))/2$.
Its sign and the leading entry $1$ give the asserted signatures.
Each conjugate pair contributes twice its Hermitian indices to $B$,
so its real indices are $2[(2,0)+(2,0)+(1,1)+(1,1)]=(12,4)$.
Both lattices span $K^2$ over $\Q$.
\end{proof}
The trace signature here refers to the normalization $G_{11}=1$.
Multiplying $G$ by a nonzero element of $K^+$ can reverse a definite
signature at a conjugate embedding, but the two indefinite signatures
persist. By absolute irreducibility every invariant Hermitian form
over $K$ is a $K^+$-multiple of $G$. Thus no such form is positive
definite at all embeddings.

\begin{proposition}[Euclidean distortion in the canonical embedding]
\label{prop:canonical-distortion}
Keep the coordinates $(u,v)$ of Example~\ref{ex:nonsimilar}.
The Euclidean norm in the canonical embedding of $K^2$ is given by
\[
 \|z\|_{\rm can}^2=\sum_{\sigma:K\hookrightarrow\C}\|\sigma(z)\|_2^2
                   =\operatorname{Tr}_{K/\Q}(z^*z).
\]
For either output lattice $L$ and any integer $m\ge 0$, let $D_L(m)$
be the supremum of $\|wz\|_{\rm can}/\|z\|_{\rm can}$ over
$0\ne z\in L$ and all words $w$ of at most $m$ letters in
$U^{\pm1},W^{\pm1}$. Put $t=2+\sqrt{2}$,
$\rho=(t+\sqrt{t^2-4})/2$ and $C=(\sqrt{4+t}+\sqrt{t})/2$.
The value of $D_L(m)$ is the same for both lattices, and
\begin{equation}\label{eq:canonical-distortion}
 \rho^{\lfloor m/4\rfloor}\ \le\ D_L(m)\ \le\ C^m,
 \qquad D_L(1)=C .
\end{equation}
\end{proposition}
\begin{proof}
For each fixed $w$, the operator norm on $K^2\otimes_{\Q}\R$ is
$\max_\sigma\|\sigma(w)\|_2$, since $w$ acts blockwise at the
complex places. This is also the supremum on either lattice:
its rational span is dense in the real space, and clearing
denominators leaves the norm ratio unchanged.

At every embedding each generator or inverse is triangular with
diagonal entries of modulus one. If its off-diagonal entry has
modulus $b$, its singular values have product one and the sum
of their squares is $2+b^2$. The larger singular value is therefore
$(\sqrt{4+b^2}+b)/2$. For $U^{\pm1}$ we have $b=1$, while for
$W^{\pm1}$ we have $b=|1-\sigma(\xi)|$, whose maximum is $\sqrt{t}$.
This gives $D_L(1)=C$ and the upper bound by submultiplicativity.

For the lower bound take $[U,W]=UWU^{-1}W^{-1}$. Multiplying the matrices
in Example~\ref{ex:nonsimilar} gives $\det[U,W]=1$ and
$\operatorname{tr}[U,W]=2-\xi-\xi^{-1}$. At $\sigma_3:s\mapsto s^3$
the trace is $t>2$, so $\sigma_3([U,W])$ has eigenvalues $\rho,\rho^{-1}$.
The word $[U,W]^j$ has a representative of length $4j$ and its operator
norm is at least $\rho^j$. Take $j=\lfloor m/4\rfloor$.
\end{proof}
Thus a perturbation $e$ of a lattice vector satisfies
$C^{-m}\|e\|_{\rm can}\le\|we\|_{\rm can}\le C^m\|e\|_{\rm can}$,
using the upper bound also for $w^{-1}$. In particular, this bounds the
amplification of a rounding error introduced before applying the word.
At the gate embedding every
word preserves the positive middle form, yet the powers above expand
exponentially in the norm that includes all embeddings. The eigenvalue
$\rho>1$ at $\sigma_3$ proves directly that the commutator has infinite
order and explains why no invariant form can be positive definite at
every embedding.

\begin{example}[An order-four relation selects a positive two-qubit space]
\label{ex:fourgate}
Keep $K=\Q(s)$, $s=\zeta_{16}$, $\lambda=\xi=s^2$ and
$\mathcal O=\Z[s]$. Write $i=\xi^2$, $\delta=1-s$ and
$\mathfrak p=\delta\mathcal O$. On $Y=K^2$ with its standard
Hermitian form take $A_1=\xi P_{12}$ and $A_2=\diag(\xi,i)$, where
$P_{12}$ interchanges the coordinates. Adjoin an orthogonal line $Z$
on which both meridians act by $-1$. The operators on $Y\perp Z$
preserve the full lattice
\[
 A=\{(x,y,w)\in\mathcal O^3:x+y+w\equiv0\pmod{\mathfrak p}\}.
\]
Indeed, modulo $\mathfrak p$ the functional $(x,y)\mapsto x+y$ on
$Y$ is invariant and the sign character on $Z$ is trivial. Their
common one-dimensional quotient over $\mathbb F_2$ gives the
congruence defining $A$. In characteristic zero the two eigenlines
of $A_2$ are exchanged by $A_1$, so $Y$ is absolutely irreducible.

Impose the single output relation $T_1^4=I$, giving a marked action
of $C_4*\Z$. The ordinary KLM representation of $Y$ acts on all of $Y^2$. Both
local kernels vanish, and $P=\lambda A_1A_2$ satisfies $P^2=\xi^7 I$,
so the kernel at infinity also vanishes.
In the ordered coordinates of $Y\oplus Y$, write $U_j=T_j|_{Y^2}$ for the two induced meridians. They are
\[
 U_1=\begin{pmatrix}0&i&\xi-1&0\\ i&0&0&i-1\\0&0&1&0\\0&0&0&1\end{pmatrix},
 \qquad
 U_2=\begin{pmatrix}1&0&0&0\\0&1&0&0\\-\xi&i&i&0\\i&-\xi&0&\xi^3\end{pmatrix}.
\]
They have orders four and eight: their respective moving blocks
have eigenvalues $\{i,-i\}$ and $\{i,\xi^3\}$, all distinct from
one. To see how the relation selects this four-dimensional space,
let $\Delta_4=T_1^4-I$ on the six-dimensional space $\KLM_\lambda(Y\oplus Z)$. It vanishes
on $Y^2$. For the standard basis $f_1,f_2$ of the sign summand $Z^2$,
$T_1f_1=-\xi f_1$ and $T_1$ acts trivially modulo $Kf_1$. Hence
\[
       \Delta_4f_1=-2f_1,\qquad \im \Delta_4=Kf_1,\qquad \dim_K\ker \Delta_4=5.
\]
For the quotient to carry both meridians, its kernel must contain
the invariant span $J$ of $\im \Delta_4$. The second meridian gives
\[
       T_2 f_1=f_1-2\xi f_2,\qquad
       J=\operatorname{span}_K\{f_1,T_2f_1\}=Z^2.
\]
Thus the quotient by $J$ is precisely $Y^2$, in agreement with
Theorem~\ref{th:transfer}. Its marked action is absolutely
irreducible by \cite[Cor.~5.15]{HN}.

For an explicit Hermitian matrix in these coordinates put
\[
 \begin{aligned}
 M_{11}&=(1+\xi)I-(\xi^{-1}+\xi^2)P_{12},\qquad
 M_{12}=(\xi^{-1}P_{12}-I)(A_2-I),\\
 M_{22}&=(A_2^{-1}-\xi I)(A_2-I),\qquad
 H=\begin{pmatrix}s^{-1}M_{11}&s^{-1}M_{12}\\sM_{12}^*&s^{-1}M_{22}\end{pmatrix}.
 \end{aligned}
\]
This is the block formula of Section~\ref{sec:tangent}. The following
table gives the eigenangle sums in units of $1/16$ for the embeddings
$s\mapsto s^j$. The square root for the signature formula is $s^j$,
and the spectrum of the conjugate of $P$ is $\{s^{7j},-s^{7j}\}$.
\[
\begin{array}{c|rrrr|c}
 j & 16\ell\dim Y & 16\theta(A_1) &16\theta(A_2)&16\theta(P)&
       \operatorname{sig}(H^{(j)})\\\hline
 1&4&12&6&22&(4,0)\\
 3&12&20&18&18&(2,2)
\end{array}
\]
The blocks $M_{11},M_{12},M_{22}$ depend only on $s^2$, so replacing $s^j$ by
$-s^j$ negates $H^{(j)}$ and exchanges its two indices. Complex
conjugation gives $H^{(-j)}=\overline{H^{(j)}}$ with the same
signature (indices $j$ modulo $16$).
Thus $j=1,15$ give $(4,0)$, $j=7,9$ give $(0,4)$, and
$j=3,5,11,13$ give $(2,2)$.
One can also read these signatures directly from $H$. Put
$c_r=s^r+s^{-r}$. In the displayed basis the four leading principal
minors are $c_1,\ 2c_2,\ 2c_3,\ 12-8c_2$, respectively.
Since $c_2^2=2$, the last minor is positive at every embedding.
The number of sign changes in the sequence starting with $1$
and followed by the four minors gives the negative index.
At $j=1$, the middle space of the sign character has signature $(1,1)$,
since both local angles are $1/2$ and $\lambda(-1)^2=\lambda$.
The orthogonal sum therefore has signature $(5,1)=(4,0)+(1,1)$.
Passing to the quotient by $J=Z^2$ removes one positive and one
negative direction. Choosing an orthonormal basis for the resulting
positive definite space identifies it with a two-qubit register as
in Proposition~\ref{prop:gateencoding}.

Write $\Lambda'=\{(x,y)\in\mathcal O^2:x+y\in\mathfrak p\}$.
Then $\pi_Y A=\mathcal O^2$, by taking $w=-x-y$, and
$A\cap Y=\Lambda'$. As there is no middle kernel on $Y^2$, the
two stable lattices and their quotient are
\[
       L_{\rm proj}=\mathcal O^4,\qquad
       L_{\rm int}=\Lambda'\oplus\Lambda',\qquad
       L_{\rm proj}/L_{\rm int}\simeq\mathbb F_2^2 .
\]
In fact the projection $A\to\mathcal O^2$ has no equivariant
$\mathcal O$-linear section. Over $K$ every equivariant section is
$y\mapsto(y,0)$, since there is no nonzero intertwiner from the
irreducible two-dimensional input $Y$ to the character $Z$.
This section sends $(1,0)$ outside $A$.
The basis $(1,-1),(0,\delta)$ of each copy of $\Lambda'$ shows
that the determinant of any linear map $F$ taking $L_{\rm proj}$
onto $L_{\rm int}$ is $\delta^2$ times a unit of $\mathcal O$.
If $F^*HF=bH$ for $b>0$, then $b\in K^+=K\cap\R$ and
$b^4=\det F\,\overline{\det F}$. Taking norms gives
$N_{K/\Q}(b)^4=16$, hence $N_{K/\Q}(b)=2$. This contradicts
$N_{K/\Q}(b)=N_{K^+/\Q}(b)^2$. Thus these Hermitian lattices
are nonsimilar even after forgetting the group action.
They are also nonisomorphic as marked modules. Any intertwiner would
extend to a scalar by absolute irreducibility. Each coordinate
projection of $L_{\rm int}$ is all of $\mathcal O$, so such a scalar
would be a unit and could not give a proper sublattice.
\end{example}

\begin{proposition}[Projective density of the two finite-order gates]
\label{prop:fourclosure}
For $U_1,U_2,H$ of Example~\ref{ex:fourgate}, the closure in the ordinary
topology is
\[
 \overline{\langle U_1,U_2\rangle}
       =\{g\in U(H):\det g\in\mu_8\}.
\]
Here $\mu_8$ is the group of eighth roots of unity. In an orthonormal
basis for $H$, the closure is $\{g\in U(4):\det g\in\mu_8\}$.
Thus words in the two gates approximate every two-qubit unitary on
this register up to a global phase. In every unitary identification
with two qubits, at least one of the two gates takes a product state
to an entangled state.
\end{proposition}
\begin{proof}
The image is infinite. Otherwise the conjugate representation at
$s\mapsto s^3$ would preserve a positive definite metric by averaging.
The same eigenline argument gives absolute irreducibility of the
conjugate input, and KLM preserves irreducibility. By Schur's lemma,
the conjugate KLM form would then be a real multiple of the positive
definite form obtained by averaging. Its signature is $(2,2)$, a
contradiction.

We apply the adjoint-representation criterion for density of
Sawicki and Karnas \cite[Lem.~3.4]{SK} to the projective image in
$PU(4)$. In this case it suffices to verify irreducibility on
$\mathfrak{sl}_4(\C)$. We give an exact certificate for the two
matrices in Example~\ref{ex:fourgate}. Put
$X=e_4(-\xi,i,i-1,0)$, where $e_4$ is the fourth coordinate column.
Then $U_2 X U_2^{-1}=\xi X$. The $\xi$-eigenspace for conjugation by $U_2$ is one-dimensional.
Indeed, among the eigenvalues $1,1,i,\xi^3$ of $U_2$, the only ordered
ratio equal to $\xi$ is $\xi^3/i$, and both eigenvalues in this ratio
are simple.

Use $1,2$ as letters for $U_1,U_2$ in the usual product order and let
$\varepsilon$ denote the empty word. For the following words $v$,
in the order shown, place the row-major entries of $vXv^{-1}$,
omitting the last diagonal entry, in the rows of a $15\times15$ matrix:
\[
 \begin{gathered}
 \varepsilon,1,11,21,111,211,121,221,\\
 2111,1211,2211,1121,2121,1221,2221 .
 \end{gathered}
\]
Under $\mathcal O\to\mathbb F_{17}$, $s\mapsto3$, its determinant
is $10\bmod17$. This computation uses the two displayed meridians
and is reproduced in \path{anc/verify_marked_relations.py}.
The omitted entry is determined by the trace. Hence the orbit of
$X$ spans $\mathfrak{sl}_4(\C)$. The conjugation representation
is unitary for the Hilbert--Schmidt inner product induced by $H$.
The orthogonal projection onto any invariant subspace commutes with
both generators and acts on the simple $\xi$-eigenline by zero or one.
Since $X$ is cyclic, that projection is zero or the identity.

Let $G$ be the compact closure. Since $\det U_1=1$ and $\det U_2=\xi^5$,
$\det(G)=\mu_8$. Hence $G^\circ\subseteq SU(H)$ by connectedness.
By the closed subgroup theorem $G$ is a compact Lie group, of positive
dimension since it is infinite. Its complexified Lie algebra is thus a
nonzero subspace of $\mathfrak{sl}_4(\C)$, invariant under conjugation
by $G$. Adjoint irreducibility gives equality, so $G^\circ=SU(H)$.
The determinant image now gives the asserted closure.

If both gates preserved product states, so would their inverses,
since both gates have finite order. The group and its closure would then
preserve the closed set of product lines in $\mathbb P^3$. This is
impossible since $PU(4)$ acts transitively on $\mathbb P^3$ and
the product lines form a nonempty proper subset.
\end{proof}

\subsection{Two phase reflections with prescribed overlap}
\label{sec:prescribed-pair}

The phase and the angle between two moving lines determine their ordered
pair of unitary reflections up to simultaneous unitary conjugacy.
We realize the same pair in two positive rank-four systems with different
local monodromy at infinity. Write $e(u)=\exp(2\pi i u)$.

\begin{proposition}\label{prop:prescribed-pair}
Let $0<t\leq 1/4$ and $0<c<1$. Set
\[
 \alpha=\pi^{-1}\arcsin(c\sin\pi t),\qquad
 \ell=t-\alpha,\qquad a=e(\alpha),\qquad \lambda=e(\ell).
\]
Let $X$ exchange the two coordinates of $\C^2$ and let $D=\diag(a,1)$.
The marked tuple $(X,-X,D,D)$ is irreducible. Its middle convolution
at $\lambda$ has a positive definite form of rank four. The third and
fourth meridians have moving eigenvalue $e(t)$, and unit vectors
$v_3,v_4$ on their moving lines satisfy $|\langle v_3,v_4\rangle|=c$.
The eigenvalue multiplicities at infinity are $(2,1,1)$.
\end{proposition}
\begin{proof}
The two eigenlines of $D$ are exchanged by $X$. The four input moving
ranks sum to four. The input product is $-D^2$, and the eigenangles
in turns of $-\lambda D^2$ are $1/2+\ell$ and $1/2+\ell+2\alpha$.
Both lie in $(0,1)$ since $\ell+2\alpha=t+\alpha<1/2$. Thus the infinity
kernel vanishes. For the square root $e(\ell/2)$, the negative index in
the signature formula of \cite[Cor.~6.9]{Neg} is
$2\ell+(1+2\alpha)-(1+2\ell+2\alpha)=0$.
After removing the local fixed lines, the last two coordinates are the
moving lines of the last two meridians, with eigenvalue $\lambda a=e(t)$.
The diagonal entries of the Hermitian form on these coordinates are
$e(-\ell/2)(a^{-1}-\lambda)(a-1)$ and the off-diagonal entry has
modulus $|a-1|^2$. Normalization gives
\[
 |\langle v_3,v_4\rangle|
 =\frac{|a-1|}{|a^{-1}-\lambda|}
 =\frac{\sin(\pi\alpha)}{\sin(\pi t)}=c .
\]
For the spectrum at infinity, write $g=(X,-X,D,D)$,
$B(y)=\sum_j(g_j-I)y_j$ and $(Ax)_j=g_{j+1}\cdots g_4x$, with the
empty product equal to $I$. Multiplication of the convolution block rows
gives the finite product $\lambda(I+AB)$, while telescoping gives
$BA=-D^2-I$. The maps descend to the quotient by the local fixed lines.
Since $BA$ is invertible, the middle space splits as $\ker B\oplus\im A$,
with both summands of dimension two. The finite product has eigenvalues
$\lambda,\lambda,-\lambda,-\lambda a^2$; inversion preserves their
multiplicities.
\end{proof}

For comparison, take Haraoka's rank-four Pochhammer system with
$\lambda_1=\cdots=\lambda_4=t$ and $\rho=\ell$.
For rational $t,\alpha$ the hypotheses of
\cite[Thm.~1.1 and Props.~1.4--1.5]{HarPoch} hold since
$0<\rho<t$ and $0<4t-3\rho=t+3\alpha<1$.
The generators have the same moving eigenvalue $e(t)$, and normalizing
the off-diagonal entry of the Gram matrix (1.15) in \cite{HarPoch}
by the common diagonal entry gives overlap
$|1-e(t-\rho)|/|1-e(t)|=\sin(\pi\alpha)/\sin(\pi t)=c$. For real parameters in this range the
rank-one form comparison in \cite[Prop.~7.1 and Cor.~6.9]{Neg} gives
the same conclusion. The Pochhammer exponents at infinity
$-\rho,-\rho,-\rho,-(4t-3\rho)$ give multiplicities $(3,1)$.
Thus the two selected operators agree in a common unitary encoding,
whereas the infinity spectra distinguish the full marked systems.
The spectral type $31,31,31,31,211$ and its reduction to rank two
also occur in \cite[Ch.~10, p.~108]{OshWeyl}.

At $t=1/4$ and $\alpha=1/6$ one has $c=1/\sqrt2$ and $\ell=1/12$.
All four input matrices have finite order and preserve the standard
cyclotomic lattice $\mathcal O^2$ for $\mathcal O=\Z[\zeta_{12}]$.
Since $\lambda=\zeta_{12}$ is a unit, the image of
$(\mathcal O^2)^4$ in the middle quotient is a full rank-four
$\mathcal O$-lattice preserved by all four output meridians and their
inverses. The two selected reflections fix the orthogonal complement
of their two moving lines pointwise. Their subgroup is therefore
projectively nondense in rank four. Projective density for the group
generated by all four meridians at this specialization remains to be
determined. After rephasing one vector, $(v_4,v_3)$ has the same
Gram matrix as $(|11\rangle,|\Phi^+\rangle)$, where
$|\Phi^+\rangle=(|00\rangle+|11\rangle)/\sqrt2$. The isometry on their
span extends to one encoding of the whole register. In this basis the
fourth meridian is controlled-$S$ and the third is
$I+(i-1)|\Phi^+\rangle\langle\Phi^+|$. Conjugating the third by the
local Pauli $I\otimes Y$ changes its moving line to the singlet. Its
square then acts as $-1$ on the singlet and as $1$ on the symmetric
subspace, hence equals the swap operator.

\section{Functions and analytic realizations}\label{sec:analytic}
\subsection{Character functions on a cyclic cover of an elliptic curve}
\begin{definition}[Middle convolution of a marked continuation space]\label{def:continuation}
Let $X$ be a connected complex manifold, and suppose that the
continuations of a multivalued meromorphic function span a
finite-dimensional space of germs at a regular base point. Assume that this space is the complexification of a representation
$V$ of $\pi_1(X)$ with a stable full lattice and an invariant
Hermitian form that is positive definite at the chosen embedding.
Fix a presentation by ordered meridians (more generally, marked
generators) $\pi_1(X)=F_n/N$. With the parameter and square root
of Section~\ref{sec:quot}, the middle convolution of this marked
continuation space is the representation of Theorem~\ref{th:ideal}, taking
$B=1$ in the absence of prescribed braid symmetries.
\end{definition}

The resulting representation determines a local system on $X$. Changing the basis of the input representation gives an isomorphic
local system. To realize its local sections as germs of scalar
functions, one must specify an additional identification.

\begin{proposition}[An elliptic character family]\label{th:ellipticchars}
Fix $\operatorname{Im}\tau>0$, $m\ge3$, $a=e^{2\pi i/m}$ and
$\Lambda=\Z+\Z\tau$. Let $\Lambda'$ be the kernel of
$\Lambda\to\Z/m$, $u+v\tau\mapsto u+v\bmod m$.
On the cyclic cover $\C/\Lambda'\to\C/\Lambda$ form the character sums
\[
        f_a(z)=\sum_{j=0}^{m-1}a^{-j}\wp_{\Lambda'}(z+j).
\]
Mark the periods $1,\tau$. With input lattice $\mathcal O f_a$,
$\mathcal O=\Z[a]$, and input normalization $\|f_a\|=1$, the
middle convolution for $\lambda=a^{-2}$ and $s=-a^{-1}$ has
continuation character $(a^{-1},a^{-1})$. The corresponding lattice is $\mathcal O f_{a^{-1}}$, and its
generator $f_{a^{-1}}$ has squared length $|1-a|^2$.
In this normalization the identification sends the class of
$(u f_a,v f_a)$ to $(u-av) f_{a^{-1}}$, and $f_{a^{-1}}(z)=f_a(-z)$.
\end{proposition}
\begin{proof}
The classes of $0,\ldots,m-1$ modulo $\Lambda'$ are distinct.
The double pole of the $j=0$ summand at zero therefore gives
$f_a\ne0$. Reindexing the sum gives $f_a(z+1)=a f_a(z)$.
Since $\tau-1\in\Lambda'$, the same holds for translation by $\tau$.
The function extends meromorphically across its poles, so its local
monodromy around each pole is trivial. Evenness and periodicity of $\wp$ give $f_a(-z)=f_{a^{-1}}(z)$.

Both input matrices are $a$. Substitution in the two KLM rows gives
\[
 T_1=\begin{pmatrix}a^{-1}&a-1\\0&1\end{pmatrix},
 \qquad
 T_2=\begin{pmatrix}1&0\\a^{-2}(a-1)&a^{-1}\end{pmatrix}.
\]
For $\eta=(1,-a)$ the Hermitian matrix is $|1-a|^2 \eta^*\eta$ and
$\eta T_1=\eta T_2=a^{-1}\eta$. Thus the radical is $\ker\eta$, and the two induced operators on the
entire middle space satisfy the commutator relation of
$\pi_1(\C/\Lambda)$.
Since $\eta(\mathcal O^2)=\mathcal O$, these identities also prove the
claims about the lattice and metric. The character sum supplies the
stated scalar function model for the descended representation.
\end{proof}

\subsection{An Euler integral for the cyclic-to-dihedral transformation}
\begin{proposition}\label{prop:dihedralperiod}
Put $\nu=1/m$, $m\ge3$, $p=-1/2-\nu$, and take positive branches
on $0<t<z<1$. The continuations of
\[
 F(z)=\int_0^z t^p(1-t)^p(z-t)^\nu\,dt
\]
span the rank-two solution space of
\[
     z(1-z)F''+(1/2-z)F'+\nu^2 F=0 .
\]
For meridians about $0$ and $1$, the marked monodromy is simultaneously
conjugate to the two KLM meridians in Example~\ref{ex:dihedral}.
\end{proposition}
\begin{proof}
The function $t^p(1-t)^p$ has local monodromy at both $0$ and $1$ equal to
$e^{2\pi i p}=-e^{-2\pi i\nu}$ and the Kummer factor has character
$e^{2\pi i\nu}$, exactly the parameters of that example.
The endpoint integral converges since $p>-1$ and $\nu>0$.
Substituting $t=zu$ gives
\[
 F(z)=\mathrm B(p+1,\nu+1)z^{1/2}
 {}_2F_1(1/2+\nu,1/2-\nu;3/2;z).
\]
The Gauss equation for this expression is the one displayed above.
On writing $z=\sin^2\phi$ it becomes $F_{\phi\phi}+4\nu^2F=0$.
Matching the leading term at zero gives the explicit period
\[
   F(z)=\frac{\mathrm B(p+1,\nu+1)}{2\nu}
              \sin\bigl(2\nu\arcsin\sqrt z\bigr).
\]
The two based meridians along the interval act on $\phi$ by
$\phi\mapsto-\phi$ and $\phi\mapsto\pi-\phi$. Continuation around the second meridian gives a nonzero cosine term
since $\sin(2\pi/m)\ne0$.
The two branches therefore span the full solution space.

In the basis $e^{\pm 2i\nu\phi}$ the two meridians are involutions
and their product has the distinct eigenvalues $e^{\pm2\pi i/m}$.
These data determine the ordered pair up to simultaneous conjugacy.
Indeed, for involutions $A,B$ with $R=AB$, one has $ARA=R^{-1}$.
Choose a nonzero $v$ in the $e^{2\pi i/m}$-eigenline of $R$.
Then $Av$ lies in the inverse eigenline. In the basis $(v,Av)$,
$R$ is diagonal with the eigenvalues in this order, $A$ exchanges
the two basis vectors, and $B=AR$.
The same description applies to the ordered KLM pair. The Gauss family is classical \cite{Vid}. The identification above
matches its ordered pair of monodromy operators with the KLM pair in
Example~\ref{ex:dihedral}.
\end{proof}

\paragraph{The double cover ordering two marked points.}
Proposition~\ref{prop:dihedralperiod} makes the permutation of the
two character lines visible. Put $u=e^{2i\phi}$. Then
$z=(2-u-u^{-1})/4$, the deck involution exchanges $u$ and $u^{-1}$,
and the solutions on this double cover are locally $u^{1/m}$ and
$u^{-1/m}$. A loop around $u=0$ acts on these lines by inverse
characters. The two lines and their exchange give the dihedral action.

A comparable ordering cover occurs in Das's proof of the analytic
connectedness principle \cite[proof of Theorem 3.4, Steps 4--5]{Das}.
On an intermediate model the general marked fiber in the disconnected
case is $(\mathbb P^1,p_1+p_2)$. In the case of one horizontal
component of degree two, its normalization and Stein factorization
give a double cover of the intermediate base. Over the smooth locus where the cover is unramified, its two sheets
correspond to the two orderings of the marked points. In a local
coordinate with marks $0,\infty$, an interchange acts by $u\mapsto c/u$
for some $c\ne0$. It inverts the generator of
$\pi_1(\mathbb C^\times)$ and exchanges a fiber character with its
inverse. This last character calculation explains the comparison.
The global rank-two local system and its Euler integral above are
specific to Proposition~\ref{prop:dihedralperiod}.

\subsection{An Appell family and its two signatures}\label{sec:appell}
Let $z_1=0,z_2=1,z_3=y$, and let $x$ be a fourth finite point. Assume
the four points are distinct. Take $0<a<1$ and consider the multivalued
function
\[
 u(t)=(t-x)^a\prod_{j=1}^3(t-z_j)^a.
\]
Periods are taken along a locally transported closed twisted cycle,
with dual coefficients for the chosen branch of $u$. For
$I_j=\int u(t)\,dt/(t-z_j)$ and $I_x=\int u(t)\,dt/(t-x)$, the exact
differential $du$ gives $I_x+\sum_j I_j=0$. Partial fractions give
\[
 \frac{\partial I_j}{\partial x}
    =\frac{a(I_j-I_x)}{x-z_j}
    =\frac{a(I_j+\sum_k I_k)}{x-z_j}.
\]
Thus the three rows are exactly \eqref{eq:row} with $B_j=a$,
$\mu=a$ and $c=-a$. In particular
\begin{equation}\label{eq:appH}
 H_{-a}=a(I_3+\mathbf1\mathbf1^t),\qquad
 \operatorname{Spec}(H_{-a})=\{a,a,4a\}.
\end{equation}

The scalar periods of $u(t)\,dt$ have an Euler integral representation
in terms of Appell's $F_1$. For example, take $x,y>1$ with $x\ne y$,
fix the branches, and integrate over $0<t<1$. Up to the factors $(xy)^a$, a branch
constant and the beta factor $\mathrm B(a+1,a+1)$, this period is
\[
 F_1\!\left(a+1;-a,-a;2a+2;\frac1x,\frac1y\right).
\]
The integral converges at both endpoints since $a>0$. As the
parameters vary, the twisted cycles can be transported locally. For the intersection-theoretic
description of the Lauricella $F_D$ system, including its two-variable
case $F_1$, see Matsumoto \cite{Mat}.

We apply the generalized Euler integral construction of
Gel\'fand, Kapranov and Zelevinsky
\cite[Def.~1.2, \S2.5 and Thm.~2.7]{GKZ} before fixing the
linear factors. For the Euler and box operators see also
\cite[\S6.3]{MO}.
For a locally constant closed twisted cycle take
\[
 \Phi(p,q)=\int t^a\prod_{j=1}^3(p_j+q_jt)^a\,dt .
\]
Work on the locus where the roots, $0$ and $\infty$ are distinct.
The six columns of the GKZ matrix, indexed by $p_j,q_j$, are
$(e_j,0),(e_j,1)\in\Z^3\oplus\Z$. The three factors have exponent
$a$, and the measure $t^a\,dt$ contributes the last parameter
$-a-1$, giving the parameter vector $(a,a,a,-a-1)$.
Differentiate under the integral
and integrate $d(t^{a+1}\prod_j(p_j+q_jt)^a)$ around the cycle to obtain
\begin{align*}
 (p_j\partial_{p_j}+q_j\partial_{q_j})\Phi&=a\Phi\quad(1\le j\le3),&
 \sum_j q_j\partial_{q_j}\Phi&=-(a+1)\Phi,\\
 (\partial_{p_i}\partial_{q_j}-\partial_{q_i}\partial_{p_j})\Phi&=0
 \quad(i\ne j).
\end{align*}
In the polynomial ring on the six differential symbols, the toric
ideal of this configuration is generated by the three minors
$\partial_{p_i}\partial_{q_j}-\partial_{q_i}\partial_{p_j}$ with $i<j$.
Indeed, two monomials of the same multidegree have the same degree in
each pair $(\partial_{p_j},\partial_{q_j})$ and the same total number
of $\partial_q$ factors. Exchanging a $\partial_q$ in a column with
surplus for a $\partial_p$ in a column with deficit connects the two
monomials by these minors.
The slice $(p_j,q_j)=(1,-1),(1,-1/x),(1,-1/y)$ recovers the above
integrand up to the elementary factors. This slice therefore realizes the Appell example within the GKZ
framework. For other multivariable hypergeometric systems related to
middle convolution, see \cite{MO}.

\begin{proposition}\label{prop:app}
Use the sign normalization of the multiplicative KLM form in
\cite[Thm.~1.1]{Neg}. Set $k=\lfloor4a\rfloor$ and
$r=\mathbf1_{\{4a\in\Z\}}$. If the three input monodromies and the Kummer monodromy are all equal
to $\exp(2\pi ia)$, then the signature of the multiplicative middle
form is
\[
                  (3-k,\ k-r).
\]
The additive coefficient signature in \eqref{eq:appH} is $(3,0)$
throughout $0<a<1$.
\end{proposition}
\begin{proof}
All finite local characters are nontrivial. The product of the three
input monodromies with the Kummer character is $\exp(2\pi i\,4a)$.
Hence the multiplicative local kernel is zero and the infinity kernel
has dimension $r$. In the rank-one signature formula the integer $\kappa$ of Theorem~\ref{th:ideal} is $a+3a-\{4a\}=k$, where $\{\ \}$ denotes fractional
part. The positive and negative indices are $3-k$ and $k-r$,
respectively. This also gives the signature of the twisted middle
intersection form up to its orientation convention. When $4a\notin\Z$,
the five weights $(a,a,a,a,\{-4a\})$ sum to $k+1$. When $4a\in\Z$,
the local monodromy at infinity is trivial. Filling in that puncture
leaves four weights with sum $k$. The two indices for $s$ nontrivial
weights summing to $d$ are $d-1$ and $s-d-1$, up to order, by the
rank-one Hodge dimension calculation \cite[Cor.~2.21]{DM}. Equation
\eqref{eq:appH} gives the additive assertion.
\end{proof}

\begin{center}
\begin{tabular}{@{}lccc@{}}
\toprule
 $a$ & coefficient signature & KLM signature & KLM dimension\\
\midrule
 $1/8$ & $(3,0)$ & $(3,0)$ & $3$\\
 $1/4$ & $(3,0)$ & $(2,0)$ & $2$\\
 $1/3$ & $(3,0)$ & $(2,1)$ & $3$\\
\bottomrule
\end{tabular}
\end{center}
At $a=1/3$ the forms have different inertia even up to an overall sign.
At $a=1/4$, the multiplicative middle image drops in rank, while
the differential system above still has three rows.
Theorem~\ref{th:tangent} compares forms near $t=0$ in an exponential
deformation. Here the monodromy parameters are evaluated at
$e^{2\pi ia}$.

\paragraph{Remark (the Picard specialization).}
At $a=1/3$ put $\omega=e^{2\pi i/3}$ and $\mathcal O_3=\Z[\omega]$.
The three input meridians and $\lambda$ all act by $\omega$ on the
one-dimensional input space. With $s=-\omega^2$ the ambient KLM Hermitian form has zero radical.
The output meridians $T_j$ have order three, so the relations $T_j^3=I$
retain the whole output lattice $\mathcal O_3^3$. These reflections
are also congruent to $I$ modulo $(1-\omega)$. The KLM Hermitian matrix is
$3G$, with unimodular normalization
\[
 G=\begin{pmatrix}1&-\omega&-\omega\\
 -\omega^2&1&-\omega\\-\omega^2&-\omega^2&1\end{pmatrix},
 \qquad \det G=-1,\qquad \operatorname{sig}(G)=(2,1).
\]
The projective image $\Gamma_x=\langle[T_1],[T_2],[T_3]\rangle$ acts
on the ball of negative lines. Its orbits there are discrete and
infinite: its matrices are integral over $\mathcal O_3$, and $(T_1T_2)^3$
is a nonidentity unipotent with $((T_1T_2)^3-I)^2=0$.

For the ordered Picard family $w^3=t(t-1)(t-y)(t-x)$, let
$\Gamma_{\rm ord}$ denote the projective monodromy group obtained by
allowing both $x$ and $y$ to vary. For this family, Dettweiler and
Wewers give five
monodromy generators and their invariant Hermitian form in parabolic
cohomology \cite[Thm.~3.2 and Rems.~3.3--3.4]{DW}. They attribute
the matrices to Picard \cite[p.~181]{Pic}. Here we identify the subgroup obtained by
moving $x$ alone and compute its cusp elements in KLM coordinates.
Shiga identifies the full ordered group in period coordinates with the
projective principal congruence group of level $(1-\omega)$
\cite[\S1, items (2)--(5), pp.~315--316]{Shi}. The three KLM meridians generate the image of the fundamental group
of the fiber over a fixed $y$. They describe the action on the
deck-character summand of the cyclic cover as $x$ moves around
$0$, $1$, and $y$ \cite{HN,NegMC}. Forgetting $x$ gives the exact sequence
\[
 1\longrightarrow\Gamma_x\longrightarrow\Gamma_{\rm ord}
 \longrightarrow\Delta(3,3,3)\longrightarrow1 .
\]
Indeed $[1:y:x]\mapsto[1:y]$ is the projection from a triple point
of the six-line arrangement. Libgober describes the four such line
pencils and the conic pencil in \cite[Ex.~5.2, p.~118]{Lib}.
On the complement, forgetting $x$ is the configuration-space
fibration \cite[Thm.~1]{FN} over the three-punctured sphere with
connected fiber. In the ordered ball quotient, each of the six mirrors has orbifold
order three, and the four cusps are omitted
\cite[Thm.~5, Cor.~31 and the paragraph following it]{Holz}. Thus $\Gamma_{\rm ord}$ is the quotient of the fundamental group of
the complement by the normal closure of the cubes of the six
meridians. After
quotienting by the fiber image, three meridians are trivial and
the other three are the peripheral loops of the base. Imposing the three remaining cube relations gives the Euclidean
triangle group $\Delta(3,3,3)$, so $\Gamma_x$
has infinite index and infinite covolume. Its limit set is the whole
boundary of the ball, since it is an infinite normal subgroup of a
lattice \cite[discussion before Cor.~4]{Sto}.

To compare kernels, write $f$ for this line-pencil epimorphism and
$K_{\rm tor}$ for the kernel of toroidal filling. Eyssidieux identifies
the toroidal quotient with the same triangle group using the pencil
of conics through the four cusps \cite[Prop.~5.22]{Eys}. In that
compactification the exceptional curves have weight one, so their
ordinary positive meridians $z_p$ generate the primitive cusp centers
and normally generate $K_{\rm tor}$ \cite[Lem.~3.1 and \S5.3.3]{Eys}.
For the line pencil centered at $p_*=[0:0:1]$, the exceptional curve
over $p_*$ is a section. At each of the other cusps, over $y=0,1,\infty$,
the exceptional curve occurs with multiplicity one in the corresponding
fiber. A transverse loop therefore gives, up to conjugacy,
\[
 f(z_{p_*})=1,\qquad f(z_{p_0})=a,\quad
 f(z_{p_1})=b,\quad f(z_{p_\infty})=c,
 \qquad a^3=b^3=c^3=abc=1 .
\]
Here $a,b,c$ have exact order three. For instance, order the finite
points as $(0,1,y,x)$ and let $z_{124}$ be the full twist on $0,1,x$
at $p_\infty$. In the coordinate basis of the KLM representation and in the reduced Burau basis
after forgetting $x$, respectively, representatives are
\[
 z_{124}=I+(-2\omega-1,\,\omega-1,\,0)^t e_3^t,\qquad
 f(z_{124})=\begin{pmatrix}\omega^2&1-\omega\\0&1\end{pmatrix}.
\]
Thus $\langle z_{124}\rangle\cap\Gamma_x=\langle z_{124}^3\rangle$,
with $z_{124}^3=(T_1T_2)^3$. The four full twists and the change to
the KLM basis are computed in \path{anc/verify_marked_relations.py}.
The images of the cusp centers normally generate $\Delta(3,3,3)$,
so $f(K_{\rm tor})=\Delta(3,3,3)$. Since $K_{\rm tor}$ is the
kernel of the conic-pencil epimorphism, the two pencils together
give a surjection onto $\Delta(3,3,3)\times\Delta(3,3,3)$.
On $M_{0,5}$, forgetting $x$ and $\infty$ gives the coordinates
$y$ and $u=y(x-1)/(x-y)$, respectively, with inverse $x=y(u-1)/(u-y)$.
In his description of the compactified Holzapfel cover, Eyssidieux
relates the blowup $\operatorname{Bl}_{3\,\mathrm{pts}}(E^2)$ to the
coordinatewise quotient of $E^2$ by the group $\mu_3^2$ of order nine,
where
$E=\C/\Z[\omega]$ \cite[\S\S5.3.2--5.3.3]{Eys}. The formulas
for $(u,y)$ above give our coordinates on the open configuration space.
For the line pencil with modified boundary weights see the proof
of \cite[Prop.~5.26, p.~18]{Eys}. For forgetful maps between ball
quotients see \cite{Der}, and for a $\Delta(2,3,6)$ quotient of
the full Eisenstein--Picard group see \cite[Prop.~6.1 and Cor.~7.2, with $n=6$]{Par}.

\section{Further questions}\label{sec:questions}
Proposition~\ref{prop:dihedralperiod} gives an Euler integral when the
input is cyclic and the output is dihedral. Which other pairs of marked
groups admit an integral transformation realizing their algebraic
middle convolution?

A proposed analytic realization must account for two features of the
algebraic construction. On the coefficient side, it must identify the
local kernels and the diagonal kernel and describe how they depend on
the chosen direction. On the monodromy side, one needs a pairing on
cycles compatible with analytic continuation and, when labels move,
with the induced permutations of the parameters. In the Appell family
one can compute both forms explicitly and see where their signatures
differ.

For $q$-difference systems a concrete question is to construct a
pairing between suitable Jackson cycles whose radical maps to
$\ker H_{1-q^{\lambda_q}}$, and then determine the transported metric.
For a wider GKZ class the analogous task is to find an explicit map
from convolution coordinates to twisted cycles and compare their
Hermitian matrices and resonance loci. For marked groups, the distinction
in Proposition~\ref{prop:fill} raises a further question: which
additional duality and action data identify the peripheral homology
quotient with the output of Theorem~\ref{th:descent}? A related
question is whether any larger class of changes of marking yields
canonical isometries beyond the chosen braid symmetries. For a family of rational curves with two marked points as in
\cite{Das}, which finite-order characters of the punctured fiber
extend over the ordered family? When do a character and its inverse
together descend to a rank-two local system on the unordered family?

For discrete Painlev\'e equations, a particular transformation of
parameters offers a starting point for comparison. Sakai realizes affine
Weyl symmetries by birational maps of rational surface families
\cite[\S6, p.~195]{Sak}. One such translation gives a discrete
Painlev\'e~II equation \cite[\S7, p.~201]{Sak}. For
$q$-Painlev\'e VI, Sasaki, Takagi and Takemura identify the
$q$-middle-convolution action on parameters at $\widetilde c=\chi_2$
with the word $s_5s_2s_1s_0s_2s_3s_2s_0s_1s_2$ in $W(D_5^{(1)})$
\cite[Prop.~4.1, eq.~(4.8)]{STT}. To recover this action from Hermitian
quotient data one must account for the null root $\delta$.
The affine root lattice maps to its finite root lattice modulo
$\Z\delta$, and affine Weyl translations act trivially on that
quotient. Can one construct an integral lattice before taking the
radical quotient, with a distinguished line mapping to
$\Z\delta$ in the Picard lattice, and lift the parameter action to
it? Such a comparison would distinguish the translation action on
the surface family from the action on a single solution space.

\section*{Acknowledgments}
I thank Yousuke Ohyama and Hidetaka Sakai for their talks at the Fourth
Tokushima Workshop on Classical Analysis. The workshop was held in
November 2026 to mark the previous year's centenary of the formation of
Paul Painlev\'e's second cabinet. Ohyama's talk on the
connection problem for higher-order $q$-hypergeometric equations and
Sakai's talk on linear $q$-difference equations helped shape the section
on $q$-middle convolution. I thank Mikio Nakahara for suggesting that I
illustrate the construction with an explicit quantum circuit. His suggestion
motivated Example~\ref{ex:local-residue-circuit}.

\noindent\emph{AI-assisted research and writing.}
I used ChatGPT (OpenAI) to help develop mathematical arguments, draft and
revise the manuscript, and write code to check the examples. Claude
(Anthropic) helped me review proofs and calculations and suggested the
input representation and congruence lattice in Example~\ref{ex:fourgate}.
I also used Gemini (Google) to review the wording and suggest clearer phrasing.
I take responsibility for the results and their presentation.


\begin{thebibliography}{99}
\bibitem{NegGates} H. Negami, Quantum gates from the middle convolution of twisted Burau representations, \arxiv{2610.00293} (2026). \doi{10.48550/arXiv.2610.00293}.
\bibitem{HarKZ} Y. Haraoka, Multiplicative middle convolution for KZ equations, \emph{Math. Z.} \textbf{294} (2020), 1787--1839. \doi{10.1007/s00209-019-02322-9}.
\bibitem{HN} K. Hiroe and H. Negami, Long-Moody construction of braid representations and Katz middle convolution, \arxiv{2303.05770} (2023). \doi{10.48550/arXiv.2303.05770}.
\bibitem{NegMC} H. Negami, Long-Moody construction of braid group representations and Haraoka's multiplicative middle convolution for KZ-type equations, \arxiv{2503.14840} (2025). \doi{10.48550/arXiv.2503.14840}.
\bibitem{BH} F. Beukers and G. Heckman, Monodromy for the hypergeometric function \({}_nF_{n-1}\), \emph{Invent. Math.} \textbf{95} (1989), 325--354. \doi{10.1007/BF01393900}.
\bibitem{NegFinite} H. Negami, Finite images of braid group representations and algebraic solutions of KZ-type equations, \arxiv{2609.25805} (2026). \doi{10.48550/arXiv.2609.25805}.
\bibitem{RRV} D. Radchenko and F. Rodriguez Villegas, Goursat rigid local systems of rank four, \emph{Res. Math. Sci.} \textbf{5} (2018), article 38. \doi{10.1007/s40687-018-0156-y}.
\bibitem{GV} E. Ghate and T. N. Venkataramana, Sums of fractions and finiteness of monodromy, \emph{Indag. Math. (N.S.)} \textbf{28} (2017), 1183--1199. See also Lemma~11 of \arxiv{1612.04548}.
\bibitem{Katz} N. M. Katz, \emph{Rigid Local Systems}, Annals of Mathematics Studies 139, Princeton University Press, 1996. \doi{10.1515/9781400882595}.
\bibitem{DR00} M. Dettweiler and S. Reiter, An algorithm of Katz and its application to the inverse Galois problem, \emph{J. Symbolic Comput.} \textbf{30} (2000), 761--798. \doi{10.1006/jsco.2000.0382}.
\bibitem{DRpre} M. Dettweiler and S. Reiter, On the middle convolution, \arxiv{math/0305311} (2003). \doi{10.48550/arXiv.math/0305311}.
\bibitem{DR} M. Dettweiler and S. Reiter, Middle convolution of Fuchsian systems and the construction of rigid differential systems, \emph{J. Algebra} \textbf{318} (2007), 1--24. \doi{10.1016/j.jalgebra.2007.08.029}.
\bibitem{Har} Y. Haraoka, Middle convolution for completely integrable systems with logarithmic singularities along hyperplane arrangements, \emph{Adv. Stud. Pure Math.} \textbf{62} (2012), 109--136. \doi{10.2969/aspm/06210109}.
\bibitem{Osh} T. Oshima, Middle convolutions of KZ-type equations and single-elimination tournaments, \arxiv{2504.09003} (2025). \doi{10.48550/arXiv.2504.09003}.
\bibitem{SY} H. Sakai and M. Yamaguchi, Spectral types of linear $q$-difference equations and $q$-analog of middle convolution, \emph{Int. Math. Res. Not. IMRN} (2017), 1975--2013. \doi{10.1093/imrn/rnw089}. Preprint: \arxiv{1410.3674}.
\bibitem{AT} Y. Arai and K. Takemura, Reformulation of $q$-middle convolution and applications, \emph{SIGMA} \textbf{22} (2026), 040. \doi{10.3842/SIGMA.2026.040}.
\bibitem{AT23} Y. Arai and K. Takemura, On $q$-middle convolution and $q$-hypergeometric equations, \emph{SIGMA} \textbf{19} (2023), 037. \doi{10.3842/SIGMA.2023.037}.
\bibitem{Neg} H. Negami, A closed signature formula for the Katz--Long--Moody Hermitian form, \arxiv{2609.11793} (2026). \doi{10.48550/arXiv.2609.11793}.
\bibitem{DW} M. Dettweiler and S. Wewers, Variation of parabolic cohomology and Poincar\'e duality, \emph{S\'emin. Congr.} \textbf{13} (2006), 145--164. Preprint: \doi{10.48550/arXiv.math/0411119}.
\bibitem{DS} M. Dettweiler and C. Sabbah, Hodge theory of the middle convolution, \emph{Publ. Res. Inst. Math. Sci.} \textbf{49} (2013), 761--800. \doi{10.4171/PRIMS/119}.
\bibitem{DSerr} M. Dettweiler and C. Sabbah, Erratum to Hodge theory of the middle convolution, \emph{Publ. Res. Inst. Math. Sci.} \textbf{54} (2018), 427--431. \doi{10.4171/PRIMS/54-2-8}.
\bibitem{Ari} D. Arinkin, Fourier transform and middle convolution for irregular $D$-modules, \arxiv{0808.0699} (2008). \doi{10.48550/arXiv.0808.0699}.
\bibitem{Yam} D. Yamakawa, Middle convolution and Harnad duality, \emph{Math. Ann.} \textbf{349} (2011), 215--262. \doi{10.1007/s00208-010-0517-3}.
\bibitem{Hir} K. Hiroe, Middle convolution for Lie algebra representations, \arxiv{2605.09828} (2026). \doi{10.48550/arXiv.2605.09828}.
\bibitem{Sou} A. Souli\'e, Generalized Long--Moody functors, \emph{Algebr. Geom. Topol.} \textbf{22} (2022), 1713--1788. \doi{10.2140/agt.2022.22.1713}.
\bibitem{CBS} W. Crawley-Boevey and P. Shaw, Multiplicative preprojective algebras, middle convolution and the Deligne--Simpson problem, \emph{Adv. Math.} \textbf{201} (2006), 180--208. \doi{10.1016/j.aim.2005.02.003}.
\bibitem{Boa} P. Boalch, From Klein to Painlev\'e via Fourier, Laplace and Jimbo, \emph{Proc. London Math. Soc.} (3) \textbf{90} (2005), 167--208. \doi{10.1112/S0024611504015011}.
\bibitem{DRJ} M. Dettweiler and S. Reiter, Painlev\'e equations and the middle convolution, \emph{Adv. Geom.} \textbf{7} (2007), 317--330. \doi{10.1515/ADVGEOM.2007.019}.
\bibitem{Vay} A. Vayalinkal, Enumerating finite braid group orbits on $SL_2(\C)$-character varieties, \emph{\'Epijournal G\'eom. Alg\'ebrique} \textbf{10} (2026), article 10. \doi{10.46298/epiga.2026.14609}.
\bibitem{Vid} R. Vidunas, Dihedral Gauss hypergeometric functions, \emph{Kyushu J. Math.} \textbf{65} (2011), 141--167. \doi{10.2206/kyushujm.65.141}.
\bibitem{FLW} M. H. Freedman, M. J. Larsen and Z. Wang, The two-eigenvalue problem and density of Jones representation of braid groups, \emph{Commun. Math. Phys.} \textbf{228} (2002), 177--199. \doi{10.1007/s002200200636}.
\bibitem{Lek} V. P. Leksin, Monodromy representations of Jordan--Pochhammer systems, \emph{Contemp. Math. Fundam. Dir.} \textbf{72} (2026), 94--101. \doi{10.22363/2413-3639-2026-72-1-94-101}.
\bibitem{GS} B. Giles and P. Selinger, Exact synthesis of multiqubit Clifford+$T$ circuits, \emph{Phys. Rev. A} \textbf{87} (2013), 032332. \doi{10.1103/PhysRevA.87.032332}.
\bibitem{KS} V. Kliuchnikov and S. Sch\"onnenbeck, Multi-qubit circuit synthesis and Hermitian lattices, \arxiv{2405.19302} (2024). \doi{10.48550/arXiv.2405.19302}.
\bibitem{LPR} V. Lyubashevsky, C. Peikert and O. Regev, On ideal lattices and learning with errors over rings, \emph{J. ACM} \textbf{60} (2013), no.~6, Art.~43. \doi{10.1145/2535925}.
\bibitem{AnthLattices} Anthropic, \href{https://github.com/anthropics/formal-math/blob/main/lattices/papers/01-quasi-polynomial-quantum-algorithms.pdf}{Galactic quasi-polynomial quantum algorithms for integral and module lattices}, preprint, November 2026.
\bibitem{Hay} E. V. Haynsworth, Determination of the inertia of a partitioned Hermitian matrix, \emph{Linear Algebra Appl.} \textbf{1} (1968), 73--81. \doi{10.1016/0024-3795(68)90050-5}.
\bibitem{Fox} R. H. Fox, Free differential calculus. I. Derivation in the free group ring, \emph{Ann. of Math.} \textbf{57} (1953), 547--560. \doi{10.2307/1969736}.
\bibitem{Shen} D. Shen, \emph{Hyperbolic structures on a toric arrangement complement}, Ph.D. thesis, Utrecht University, 2015. \href{https://www.math.ru.nl/~heckman/PhD\%20Dali\%20Shen.pdf}{Thesis}.
\bibitem{Hum} J. E. Humphreys, \emph{Reflection Groups and Coxeter Groups}, Cambridge Studies in Advanced Mathematics 29, Cambridge University Press, 1990. \doi{10.1017/CBO9780511623646}.
\bibitem{FN} E. Fadell and L. Neuwirth, Configuration spaces, \emph{Math. Scand.} \textbf{10} (1962), 111--118. \doi{10.7146/math.scand.a-10517}.
\bibitem{SK} A. Sawicki and K. Karnas, Universality of single-qudit gates, \emph{Ann. Henri Poincar\'e} \textbf{18} (2017), 3515--3552. \doi{10.1007/s00023-017-0604-z}.
\bibitem{HarPoch} Y. Haraoka, Finite monodromy of Pochhammer equation, \emph{Ann. Inst. Fourier} \textbf{44} (1994), 767--810. \doi{10.5802/aif.1417}.
\bibitem{OshWeyl} T. Oshima, \emph{Fractional Calculus of Weyl Algebra and Fuchsian Differential Equations}, MSJ Memoirs 28, Mathematical Society of Japan, 2012. \arxiv{1102.2792}.
\bibitem{Das} S. Das, Connectedness principle in the complex analytic setting, \arxiv{2611.34654} (2026). \doi{10.48550/arXiv.2611.34654}.
\bibitem{Mat} K. Matsumoto, Monodromy and Pfaffian of Lauricella's $F_D$ in terms of the intersection forms of twisted (co)homology groups, \emph{Kyushu J. Math.} \textbf{67} (2013), 367--387. \doi{10.2206/kyushujm.67.367}.
\bibitem{GKZ} I. M. Gel\'fand, M. M. Kapranov and A. V. Zelevinsky, Generalized Euler integrals and $A$-hypergeometric functions, \emph{Adv. Math.} \textbf{84} (1990), 255--271. \doi{10.1016/0001-8708(90)90048-R}.
\bibitem{MO} S.-J. Matsubara-Heo and T. Oshima, Generalized hypergeometric functions with several variables, \emph{Indag. Math.} \textbf{36} (2025), 507--566. \doi{10.1016/j.indag.2024.07.007}.
\bibitem{DM} P. Deligne and G. D. Mostow, Monodromy of hypergeometric functions and non-lattice integral monodromy, \emph{Publ. Math. IH\'ES} \textbf{63} (1986), 5--89. \doi{10.1007/BF02831622}.
\bibitem{Pic} É. Picard, Sur les formes quadratiques ternaires indéfinies à indéterminées conjuguées et sur les fonctions hyperfuchsiennes correspondantes, \emph{Acta Math.} \textbf{5} (1884), 121--182. \doi{10.1007/BF02421555}.
\bibitem{Shi} H. Shiga, On the representation of the Picard modular function by theta constants I--II, \emph{Publ. Res. Inst. Math. Sci.} \textbf{24} (1988), 311--360. \doi{10.2977/PRIMS/1195175031}.
\bibitem{Lib} A. Libgober, Albanese varieties of abelian covers, \emph{J. Singul.} \textbf{12} (2015), 105--123. \doi{10.5427/jsing.2015.12g}.
\bibitem{Holz} R.-P. Holzapfel, Indefinite Eisenstein lattices: a modern ball-rendevous with Poincar\'e, Picard, Hecke, Shimura, Mumford, Deligne and Hirzebruch, \emph{J. Geom. Symmetry Phys.} \textbf{47} (2018), 1--40. \doi{10.7546/jgsp-47-2018-1-40}.
\bibitem{Sto} M. Stover, Cusp and $b_1$ growth for ball quotients and maps onto $\Z$ with finitely generated kernel, \emph{Indiana Univ. Math. J.} \textbf{70} (2021), 213--233. \doi{10.1512/iumj.2021.70.8191}.
\bibitem{Eys} P. Eyssidieux, Orbifold K\"ahler groups related to arithmetic complex hyperbolic lattices, \arxiv{1805.00767} (2018). \doi{10.48550/arXiv.1805.00767}.
\bibitem{Der} M. Deraux, Forgetful maps between Deligne--Mostow ball quotients, \emph{Geom. Dedicata} \textbf{150} (2011), 377--389. \doi{10.1007/s10711-010-9511-x}.
\bibitem{Par} J. R. Parker, Cone metrics on the sphere and Livn\'e's lattices, \emph{Acta Math.} \textbf{196} (2006), 1--64. \doi{10.1007/s11511-006-0001-9}.
\bibitem{Sak} H. Sakai, Rational surfaces associated with affine root systems and geometry of the Painlev\'e equations, \emph{Commun. Math. Phys.} \textbf{220} (2001), 165--229. \doi{10.1007/s002200100446}.
\bibitem{STT} S. Sasaki, S. Takagi and K. Takemura, $q$-middle convolution and $q$-Painlev\'e equation, \emph{SIGMA} \textbf{18} (2022), 056. \doi{10.3842/SIGMA.2022.056}.
\end{thebibliography}
\end{document}